\documentclass[twocolumn,prd,aps,floatfix,superscriptaddress,longbibliography,nofootinbib,notitlepage]{revtex4-2}
\usepackage{amssymb, amsmath, amsfonts, amsthm}
\usepackage{graphicx}
\usepackage{hyperref}
\usepackage{xcolor}
\usepackage{subcaption}
\usepackage{soul}
\begin{document}

\title{Known unknowns: assessing 
%Evaluating 
the impact of instrumental calibration uncertainty on LISA science}
%Setting LISA calibration requirements\jon{Are we happy with this title?}}
% The impact of LISA calibration requirements on science

\author{Etienne Savalle}
\address{AstroParticule et Cosmologie (APC), Universit\'{e} Paris Cit\'{e}/CNRS, 75013 Paris, France} \email{etienne.savalle@apc.in2p3.fr}
\author{Jonathan Gair}
\address{Max Planck Institute for Gravitational Physics (Albert Einstein Institute), Am M\"{u}hlenberg 1, Potsdam-Golm 14476, Germany} \email{jgair@aei.mpg.de}
\author{Lorenzo Speri}
\address{Max Planck Institute for Gravitational Physics (Albert Einstein Institute), Am M\"{u}hlenberg 1, Potsdam-Golm 14476, Germany} \email{lsperi@aei.mpg.de}
\author{Stanislav Babak}
\address{AstroParticule et Cosmologie (APC), Universit\'{e} Paris Cit\'{e}/CNRS, 75013 Paris, France} \email{stas@apc.in2p3.fr}

% \hfill \vspace{.35cm} \hrule \vspace{1cm}
\begin{abstract}
The primary scientific results of the future space-based gravitational wave interferometer LISA will come from the parameter inference of a large variety of gravitational wave sources. However, the presence of calibration errors could potentially degrade the measurement precision of the system parameters. Here, we assess the impact of calibration uncertainties on parameter estimation for individual sources, focusing on massive black holes, extreme-mass-ratio inspirals (EMRIs), galactic binaries, and stellar origin black hole binaries. Using a Fisher matrix formalism, we investigate how the measurement precision of source parameters degrades as a function of the size of the assumed calibration uncertainties. If we require that parameter measurements are degraded by no more than a factor of two relative to their value in the absence of calibration error, we find that calibration errors should be smaller than a few tenths of a percent in amplitude and $10^{-3}$ in phase. 
%We impose that the source parameter measurement errors cannot be more than two times larger than their value in absence of calibration errors.
%Using this threshold with all sources, we set a $0.1\%$ accuracy requirement on the calibration errors.
We also investigate the possibility of using verification binaries and EMRIs to constrain calibration uncertainties. Verification binaries can constrain amplitude calibration uncertainties at the level of a few percent, while both source types can provide constrain phase calibration at the level of a few$\times10^{-2}$.
%The former sources can provide a measurement precision of the calibration errors of order ... and of order ... when including the Electromagnetic prior information.
%We find that Extreme Mass Ratio Inspirals could potentially constrain some of the calibration parameters using solely their gravitational wave signals....
\end{abstract}

\maketitle
%\tableofcontents
\section{Introduction}
%\responsible{Stas}
LISA is a space-based gravitational wave observatory selected by ESA to be the third large mission in the Cosmic Vision programme and will be launched around 2034. Recently it has completed phase A (mission formulation phase) and is now entering phase B1 (preliminary design), which will lead to mission adoption around 2024. LISA will consist of three identical satellites in a heliocentric orbit forming an equilateral triangle with sides of length 2.5 million km. The centre of the constellation will be trailing about 20 deg. behind the Earth. LISA will detect gravitational waves (GWs) by measuring the proper distance between free-falling test masses (located inside each spacecraft) using laser interferometry. The change in the proper distance in the LISA frame (which can be seen as transverse-traceless gauge) translates into a modulation of the frequency (or phase) of the laser light and is directly measured by the phasemeters. 

The dominant noise component in LISA measurements is the laser frequency noise. Even with frequency pre-stabilization this is expected to be several orders of magnitude higher than what is required to detect gravitational wave (GW) signals. This problem is further exacerbated by the fact that LISA's arms are not of equal length, and in fact change over time as the constellation ``breathes'' due to the shape of the orbits of the individual satellites. Fortunately, the laser frequency noise can be cancelled in post-processing. LISA will use {\it transponding interferometry}, in which incoming laser light is amplified, phase-locked and then transmitted back. The phase of the light in each of the 6 laser links (one in each direction along each arm of the constellation) will be measured on the satellites using phasemeters. These phase measurements can be delayed in time and linearly combined to form closed loops of equal optical path length. Subtracting such combinations from each other cancels the laser frequency noise to an acceptable level. This technique is called Time Delay Interferometry \cite{Tinto:2020fcc}, and for it to work we need to know the distances between the three spacecrafts accurately and we need to synchronize their on-board clocks. 

Data decimation and post-processing (that includes filtering and resampling the data during clock correction and building TDI) %\jon{mention TDI again here, otherwise the previous paragraph is unconnected?}) 
is a complex procedure which leads to a non-trivial response of the detector to the incident GW signal. Inaccurate modelling of the signal chain could potentially distort the GW signal in the data that is eventually analysed. We will refer to differences between the actual GW signal present in the data and how we model it as calibration error (or mis-calibration). The calibration process, by which the mapping between an incident physical signal and a detector output is learnt or measured, plays an important role in physical experiments. For example, in the LIGO detector what is actually measured is the size of the electrostatic force that must be applied to the end mirrors to keep the detector at a dark interference fringe. The transformation between the force (or voltage) and the incident GW strain requires a  calibration process, in which the parameters of a calibration model are constrained by injecting monochromatic signals into the detector using laser modulation and measuring the system's response \cite{LIGOScientific:2016xax}. Inevitably, the calibration cannot be perfect and leads to some residual uncertainty that impacts our ability to measure the parameters of the incident GW signal. In the case of LISA we will have more direct measurement of the GW strain and so we do not expect to require a complex calibration procedure, or to have a large calibration error. However, it is still important to estimate the possible impact of any remaining calibration errors on our ability to detect and estimate the parameters of sources in the data.  

In this analysis, we will treat the GW signal in the frequency domain and decompose the calibration error into parts acting on the amplitude and on the phase. 
%Note that it is important to separate the calibration error from the unknown noise. 
The calibration error can be seen as a biased measurement of the amplitude and/or phase, and this bias could be both time and frequency dependent. In practice we will not know this bias, but it should be possible to bound it by an envelope. Here we will place limits on the size of that envelope so that we ensure LISA science exploitation is not affected. Ignoring calibration error will lead to biases in parameter estimation, similar to the biases that can appear due to errors in waveform models~\cite{2007PhRvD..76j4018C}. In common with that case, these biases will tend to be independent of the amplitude of the gravitational wave signal and hence become proportionally more important and even limiting for loud signals. In contrast to the waveform error case, time-independent calibration errors will affect all GW signals in the same way, which will have implications for inference on populations.
%Note that the calibration error exists even in the absence of the noise and (usually) independent of the strength of GW signal, implying that it might become a limiting factor in the parameters inference for loud signals. 

We adopt a particular model for the LISA calibration uncertainties, which is similar to models previously employed in the analysis of LIGO data~\cite{Vitale:2011wu}. This model is based on a natural cubic spline defined at four frequency knots, chosen to roughly split the full LISA band into low-frequency ($0.1-1$ mHz), 
mid-frequency ($1-10$ mHz) and high-frequency ($10-1000$ mHz) ranges.  
We place a prior on the weight at each knot of the spline that is a Gaussian distribution with zero mean and a width which characterises the calibration error envelope.

We use this spline model to determine the impact of calibration uncertainties on detection and parameter estimation of LISA sources, focusing primarily on the latter. We use a Fisher matrix formalism to assess our ability to simultaneously measure the calibration uncertainty and the source parameters, including the Gaussian prior on the calibration parameters to represent our knowledge of the calibration envelope. By changing the variance of the Gaussian prior we can identify the size of uncertainty at which calibration uncertainties become limiting. To make this concrete, we define this as the point at which the parameter estimation uncertainty doubles relative to its value in the absence of calibration uncertainty. We consider all classes of resolvable source expected to be observed by LISA  ---  galactic binaries, (super-) massive black hole binaries, stellar-origin black hole binaries (SBHB) and extreme-mass-ratio inspirals (EMRIs). Galactic binaries are binary systems in the Milky Way containing two compact objects, typically white dwarfs (WDs), but also neutron stars (NSs) and black holes (BHs), with orbital periods of the order of an hour. The GW signals are nearly monochromatic, modulated by the LISA motion. Some of these systems are already known through electromagnetic observations, and are termed verification binaries. For these binaries the amplitude and phase evolution of the systems are known to some extent, albeit relatively poorly. Massive black hole binaries are the strongest sources expected in the LISA data. These signals arise from the inspiral and merger of massive black holes in the centres of galaxies, following mergers between their host galaxies. The GW signals are broadband and are in the LISA band for between a few hours and a month, depending on the masses of the binary components. Stellar-origin black hole binaries are binaries of two black holes, of the type being observed merging by ground-based gravitational wave detectors. These systems are observable in the LISA band in the early-inspiral stage, between a few and a few hundred years before merger. %\stas{We cannot say that we'll also see binaries which are hundreds years prior to merger}.
Finally, EMRIs are the inspirals of stellar-origin compact objects (typically BHs, but possible NS or WD) into massive black holes in the centres of galaxies. These arise following the capture of a compact object as a result of relaxation process in the stellar cluster surrounding the massive black hole. These sources can be observed in LISA data for up to several years, giving the potential for ultra-precise measurements of the parameters of the system by tracking the GW phase over $\sim 10^6$ cycles.

For all source types, we find that calibration uncertainties typically become important when they are comparable to the precision with which the source parameters can be determined. Amplitude calibration uncertainties become limiting, although phase uncertainties do not in most cases, since the phase evolution of individual sources is typically distinct from the slowly-varying calibration uncertainty we assume here. We find that if the amplitude calibration is at the level of $0.1\%$ and phase calibration is at the level of $10^{-3}$ then parameter estimation will not be adversely affected. The paper is organised as follows. In Section~\ref{sec:calmod} we describe the formalism we will use to set calibration errors, including the model we use for calibration uncertainties and how we can assess the impact of calibration errors on both source detection and parameter estimation. In Section~\ref{sec_SingSourceReq} we assess the impact of calibration uncertainties on parameter estimation for all resolvable LISA source types and discuss qualitatively how these results can be extrapolated to understand the implications for population inference. Finally, in Section~\ref{sec_GWcalibrators} we discuss how LISA GW sources can be used as calibrators to measure the LISA calibration uncertainty, before finishing in Section~\ref{sec_summary} with a summary of our conclusions.

\section{Setting calibration requirements}
\label{sec:calmod}
The typical assumption used in data analysis of gravitational wave (and other) detectors is that the observed data stream, $d(t)$, is a linear combination of signal and noise
$$
d(t)= s(t) + n(t).
$$
We assume a model for how the signal component of the data, $s(t) = h(t|\vec\theta)$, depends on the physical parameters, $\vec\theta$, that characterise the system and which we wish to determine. The likelihood for the observed data is then $p(d(t) | \vec\theta) = p(n(t) = d(t) - h(t|\vec\theta))$. In a gravitational wave context it is usual to further assume that the instrumental noise follows a Gaussian distribution characterised by a power spectral density, $S_h(f)$. In that case the likelihood becomes
\begin{eqnarray}
p(d(t) | \vec\theta) \propto \exp\left[-\frac{1}{2} (d(t)-h(t|\vec\theta) | d(t)-h(t|\vec\theta))\right], \nonumber \\   
\label{eq:standlike}
\end{eqnarray}
where
\begin{equation}
     (a(t) | b(t)) = 4 \mbox{Re} \int_0^\infty \frac{\tilde{a}^*(f) \tilde{b}(f)}{S_h(f)}\, {\rm d}f.
\end{equation}

In reality this model will be imperfect. The noise distribution might not be Gaussian, or the PSD might be different to that assumed. We will call this \textbf{noise modelling uncertainty}. Additionally the response might be different to what we have modelled, which can be represented by writing $\tilde{s}(f) = C(f) \tilde{h}(f|\vec\theta)$, with $C(f) \neq 1$, and then
\begin{equation}
\tilde{d}(f) = C(f) \tilde{h}(f|\vec\theta) + \tilde{n}(f).
\label{eq:calmod}
\end{equation}
We will call the existence of $C(f) \neq 1$ a \textbf{calibration error}. While both noise modelling uncertainty and calibration error can have an impact on scientific inference and must be minimized, in this paper we will focus on understanding the latter.
%Inference can be based on this model, by making assumptions about the statistical properties of the noise and about the form of $C(f)$. The noise model might not be known perfectly, which will lead to biases or less precise inference, and we will call this   While both types of mismodelling are important,  % and set limits on how much $C(f)$ can deviate from unity before calibration error starts to impact scientific inference.

The above expression is not completely general, since it assumes that the data stream depends linearly on the incident gravitational wave field and the function $C(f)$ does not depend on the properties of the gravitational wave sources in the data. Non-linearities that scale like the square of the gravitational wave amplitude will be sub-dominant, but there could be effects, for example filtering errors or the presence of data gaps, that lead to leakage between frequencies such that $\tilde{d}(f)$ depends on $h(f'|\vec\theta)$ for $f' \neq f$. Calibration errors could also arise differently in different spacecraft, which would mean the calibration errors in the TDI channels would depend on the sky locations of the sources. An additional complication arises because the LISA data stream will contain multiple sources of different types, and so in reality
\begin{align}
\tilde{h}(f) = \sum_{i=1}^m \tilde{h}_i(f|\vec\theta_i)
\end{align}
where $i$ enumerates the different sources, with parameters $\vec\theta_i$ and waveforms $\tilde{h}_i(f|\vec\theta_i)$. As these sources will occur at different times, they may be affected by slightly different calibration errors. Some of these effects can be accounted for by writing down the calibration error in a different way. For dealing with data gaps, a model similar to Eq.~(\ref{eq:calmod}) but applied to the time domain data might be more appropriate, since gaps are multiplicative filters in the time domain, whereas they are convolutions in the Fourier domain. To handle the presence of multiple sources in the data, a model that includes more than one calibration error function could be used, for example
\begin{equation}
\tilde{d}(f) = C(f) \sum_{i=1}^m C_i(f) \tilde{h}_i(f) + \tilde{n}(f).
\label{eq:calmodSS}
\end{equation}
Here $C(f)$ and $C_i(f)$ can be interpreted as the global and source-$i$ specific calibration error respectively.

Regardless of how complex the true dependence of the data stream on the signal parameters is in reality, we can always define a total calibration error $C(f| \{\vec\theta_i\})$ as the ratio of the signal-dependent part of the instrumental data stream to the model assumed for that component of the data. Doing this would introduce a dependence into $C(f)$ on the parameters of the sources in the data. In this paper we will ignore these complications, and use model~(\ref{eq:calmod}) to set limits on how much $C(f)$ can deviate from unity before calibration error starts to impact scientific inference for individual sources. These results can then be used with more complete models of the LISA data to assess if the calibration requirement is met. Provided that the modelled $C(f)$ lies within the derived calibration envelope, for all sources and for any reasonable realisation of the full population of sources in the data, the conclusion that science is unaffected should be robust. If the calibration requirement is not satisfied, using a more complex model for the calibration error might partially mitigate the problem. However, those more complex models are not needed for setting a baseline requirement which is the goal of this paper.

Eq.~(\ref{eq:calmod}) applies to a single data stream, but in reality LISA data will comprise three separate TDI data channels, (A, E, T). The response in each one can be modelled in a similar way, but an assumption must be made as to whether the calibration uncertainties are independent or correlated between different channels. There are other TDI combinations that can be constructed and it may be that certain TDI combinations can be better represented by the calibration model assumed here than others. For the purpose of this paper we will use (A, E, T) and will additionally assume that the calibration error is common to the three channels. This should be a conservative assumption as it makes the calibration error maximally degenerate with the signal waveform.
%\jon{TO CHECK. Are we assuming common calibration errors, or independent calibration errors? Are we doing this consistently for all source types?}

%For ease of exposition, in the following it should be assumed we are working with a single GW frequency series and a single calibration envelope described by model~(\ref{eq:calmod}), unless otherwise stated. The approach described carries over with minor modifications to alternative representations of the data.

Calibration errors affect our ability to identify sources in the data (see Section~\ref{sec_SNR}), and our ability to accurately measure the parameters of the sources (see Section~\ref{sec_PE}). The latter drives the LISA calibration requirement since the relatively small fluctuations in LISA event rates arising from calibration uncertainties are much smaller than intrinsic astrophysical uncertainties in the LISA source population. However, for completeness,   we will describe in the following sections how to assess the impact of calibration errors on both. Before doing that, in Section~\ref{sec_calibrationmodel}, we will first describe the calibration error model that we will use to produce later results in this paper.

\subsection{Calibration model} \label{sec_calibrationmodel}
\begin{figure*}
    \centering
    \includegraphics[width=\textwidth]{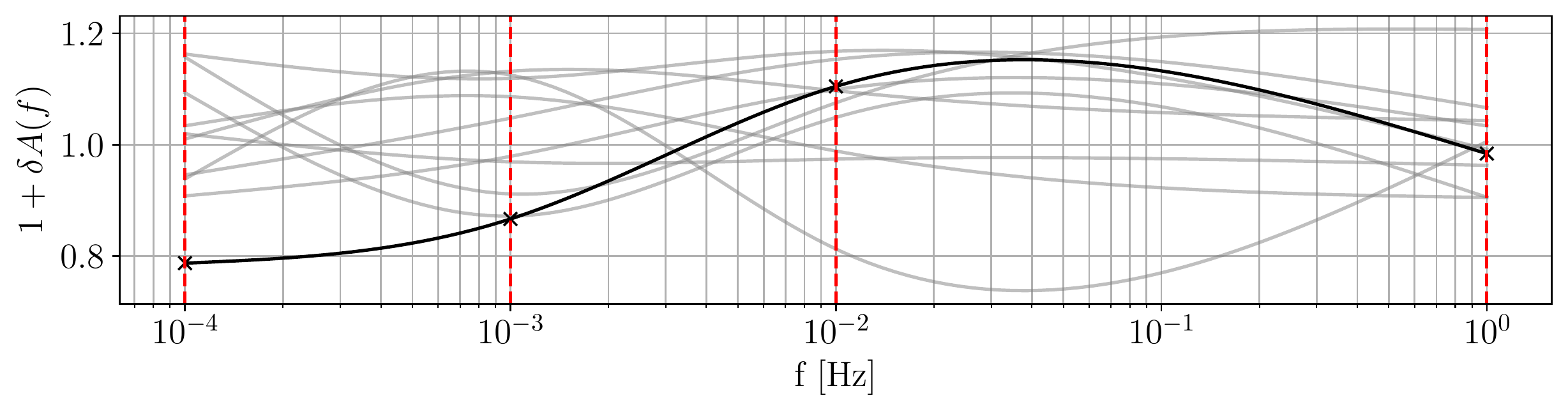}
    \caption{Example calibration envelopes. Each line is a random draw of the amplitude calibration uncertainty, $1+\delta A(f)$, assuming equal size independent uncertainties at each knot of magnitude $\sigma_C = 10^{-1}$. The vertical dashed lines show the locations of the knots. 
    %\lorenzo{make sure $\sigma_C$ is consistent with plot} \etienne{Done with $\sigma_C = 10^ {-1}$}
    }
    \label{fig:CalibEnv}
\end{figure*}

In practice, the statistical properties of the calibration error might be estimated through simulation or measurements on board the LISA satellites. Given a distribution of calibration errors these can be folded into scientific inference through, for example, reweighting of posterior samples computed ignoring calibration error~\footnote{This is the approach that is now being taken in the analysis of data from the ground-based gravitational wave detectors. See, for example~\cite{Vitale2021_LIGOCalib}. %\jon{Add ref}
}. For the purpose of setting/assessing calibration requirements we propose to take a different approach and write down a simple model for $C(f)$, the parameters of which can be interpreted as the accuracy of (amplitude and phase) calibration at different reference frequencies. First we decompose
$$
C(f) = (1+ \delta A(f)) \exp(2 \pi i \delta \phi(f)),
$$
in which $\delta A(f)$ can be interpreted as the amplitude calibration uncertainty at frequency $f$, and $\delta \phi(f)$ can be interpreted as the corresponding phase calibration uncertainty. We further assume that these uncertainties are smoothly varying over a relatively wide range of frequency. At frequencies below $1$mHz, the LISA noise is dominated by acceleration noise, while above $10$mHz it is dominated by noise in the optical metrology system. The region $1$--$10$mHz is a cross-over region~\cite{eLISA:2013xep}.  %\jon{Add ref?} 
These frequencies define the scale over which we would expect the LISA calibration error to vary. We therefore model them as natural cubic splines in $\log_{10}(f)$, with nodes at the frequencies
$$
%\{ 0, 0, 0, 0, 0, 0\}.
\{ 0.1, 1, 10, 1000\}.
$$
In this model, between any two knots ($f_i$ and $f_{i-1}$) the amplitude or phase calibration error function can be written as:
\begin{align}
\delta X(f) &= M_{i-1} \frac{\left(f_{i}-f\right)^{3}}{\left(f_i- f_{i-1}\right)}+M_{i} \frac{\left(f-f_{i-1}\right)^{3}}{\left(f_i- x_{f-1}\right)}\\
&+\left(\delta X_{i-1}-\frac{1}{6} M_{i-1} \left(f_i-f_{i-1}\right)^{2}\right) \frac{f_{i}-f}{f_i- f_{i-1}}\\
&+\left(\delta X_{i}-\frac{1}{6} M_{i} \left(f_i- x_{f-1}\right)^{2}\right) \frac{f-f_{i-1}}{f_i- f_{i-1}}
\label{eq_Cf}
\end{align}
where the four amplitude calibration parameters will be noted as $\{\delta A_0,\delta A_1,\delta A_2,\delta A_3\}$ and the four phase calibration parameters as $\{\delta \phi_0,\delta \phi_1,\delta \phi_2,\delta \phi_3\}$. The variables $M_{i}$ are unknowns that can be easily evaluated by solving a system of equations defined by the boundary conditions imposed on the interpolation function. 
%\lorenzo{notation consistent with figures } \jon{From the description in appendix B it seems like you are fitting a cubic, not a cubic spline. It won't make much difference here, but only because we have chosen to have just four knots. Cubic splines scale better as we add knots. Is this just a difference in the presentation, or an actual difference in implementation?} 
%\etienne{not relevant anymore}The relationship between the parameters $\{a_n, p_n\}$ and $\{\delta A_i, \delta\phi_i\}$  are derived in appendix \ref{sec_cubicparameters}.
To avoid calibration-error-induced biases we assume that inference on the LISA data will include the parameters describing the calibration error as additional model parameters to be marginalised over during inference. We represent our knowledge of the calibration accuracy by the priors on the spline model. For simplicity we will use Gaussian uncertainties $\sigma_C$, and assume these are independent between the amplitude and phase errors and between different frequencies. Example splines constructed in this way are shown in Figure~\ref{fig:CalibEnv}.
%(although allowing for correlations is straightforward if it is deemed appropriate).

%\jon{I made this figure for a talk and thought it would be useful to have a plot in the paper. We can discuss changes if necessary.}

%\jon{Could add more useful spline results here, e.g., an explicit expression for the spline function between each pair of knots [see GB section], or expressions for the derivatives which are analytic.}
%We will describe how calibration could be assessed with reference to a variety of different types of Figure of Merit. There are two ways that these ideas could be implemented. One approach would be to try to invert the requirements described below to directly specify constraints on the calibration uncertainties. This is possible but is complex unless simplifying assumptions are made since the calibration spline is a function of multiple parameters. The alternative approach is to modify the input to the Figures of Merit pipelines so that they take not only a specification of the noise covariance matrix, but also a specification of the calibration uncertainty. The Figures of Merit can then be evaluated accounting for the specified calibration uncertainty and given the usual red/yellow/green classification of success.

\subsection{Impact of calibration error on source detection}\label{sec_SNR}
It can be seen from Eq.~(\ref{eq:calmod}) that the effect of calibration error is to modify the effective signal component of the data. As such, it is clear that this will have some bearing on our ability to identify sources in the data. The methods by which sources will be identified in the LISA data are not yet fully specified, but for ground-based detectors source detection relies mostly on \emph{matched filtering}~\cite{PhysRevD.60.022002}. A \emph{bank of templates} is constructed, which are waveforms, $h(t|\vec\theta_i)$, for a set of parameter choices, $\vec\theta_i$, for potential sources. The waveforms are then \emph{normalised},
\begin{equation}
\hat{h}(t|\vec\theta) = \frac{h(t|\vec\theta) }{\sqrt{(h(t|\vec\theta)|h(t|\vec\theta))}},
\end{equation}
where the inner product $(a(t)|b(t))$ was defined in Eq.~(\ref{eq:standlike}).  The number of templates in the bank is chosen to ensure a sufficiently dense coverage of parameter space. The maximum \emph{signal-to-noise ratio} (SNR) across the template bank
%\qquad \mathbf{ d} = \rho_0 \mathbf{C}(\mu) \mathbf{\hat{h}}(\vec\theta_0) +\mathbf{n}
%The SNR achieved in a search is given by
$$
\hat{\rho} = \max_{\vec\theta} (d(t) | \hat{h}(t|\vec\theta)),
$$
is computed and compared to a threshold. Events above the threshold are regarded as potential gravitational wave candidates that are subsequently followed up. In the presence of calibration errors, the search would be constructed and carried out in the same way, but the observed data would be modified according to~(\ref{eq:calmod}). We assume that the observed data contains a source with true parameters $\vec\theta_0$, true signal-to-noise ratio $\rho_0^2=(h(t|\vec\theta_0)|h(t|\vec\theta_0))$ and has a true frequency-domain calibration uncertainty $C_0(f)$, so that $\tilde{d}(f) = \rho_0 C_0(f) \hat{h}(f|\vec\theta_0) + \tilde{n}$. Expanding the likelihood in $\Delta \vec\theta=\vec\theta-\vec\theta_0$, and dropping the explicit dependence of the various quantities on time or frequency for ease of notation, we obtain
%\begin{widetext}
\begin{align}
(d | \hat{h}(\vec\theta)) &=  (d  | \hat{h}(\vec\theta_0)) + (d  | \partial_i \hat{h}(\vec\theta_0)) \Delta\theta^i \nonumber \\
& \hspace{0.5in} + \frac{1}{2}  (d  | \partial_{ij} \hat{h}(\vec\theta_0)) \Delta\theta^i  \Delta\theta^j + \cdots \nonumber\\
&= \frac{1}{2} (\Delta\theta^i + D_k (H^{-1})^{ki} )H_{ij} (\Delta\theta^j +(H^{-1})^{jk} D_k) \nonumber \\
& \hspace{0.5in}+(d  | \hat{h}(\vec\theta_0)) - \frac{1}{2} D_i H_{ij}^{-1} D_j + \cdots \nonumber
\end{align}
\vspace{-0.8cm}
\begin{align}
\max_{\vec\theta}(d | \hat{h}(\vec\theta)) &=  (\rho_0 {C}_0 \hat{h}(\vec\theta_0) + n  | \hat{h}(\vec\theta_0)) \nonumber \\
&\hspace{0.5cm} - \frac{1}{2} D_i H_{ij}^{-1} D_j + \cdots \nonumber \\
\mbox{where } D_i &= (d | \partial_{i} \hat{h}(\vec\theta_0)) ), \quad H_{ij} = (d | \partial_{ij} \hat{h}(\vec\theta_0)) ).
\end{align}
%\end{widetext}
Since $(\hat{h} |\hat{h}) \equiv 1$ by construction, we deduce $(h | \partial_i\hat{h}) = 0$ and hence that the second term is quadratic in small quantities, while the first term has a linear correction in the size of the calibration uncertainty. We deduce that the expected value (i.e., the average value over realisations of the noise) of the maximum SNR in the presence of calibration uncertainties is $ (\rho_0 {C}_0 \hat{h}(\vec\theta_0) | \hat{h}(\vec\theta_0)) =  ( {C}_0 h(\vec\theta_0)  | h(\vec\theta_0))/\rho_0$, which is
\begin{equation}
\frac{1}{\sqrt{h(\vec\theta_0)|h(\vec\theta_0))}} \int_0^\infty \frac{(1 + \delta A(f)) {\rm e}^{2 \pi i \delta\phi(f)} |\tilde{h}(f|\vec\theta_0)|^2}{S_n(f)} \, {\rm d} f.
\end{equation}
The corresponding result in the absence of calibration uncertainties can be found by setting $\delta A(f) = \delta \phi(f) = 0$ in the above. The fraction of events that pass the specified SNR threshold could be evaluated using the above model combined with a population model and a model for the distribution of calibration uncertainties. However, we will not do this here for three reasons. Firstly, the modified SNR is linear in $(1+\delta A(f))$ and so will tend to average to $1$ if we use a symmetric amplitude calibration uncertainty model. This does not mean the number of detected events will be unchanged, as there are more events further away, but it will partially mitigate the impact. Secondly, it is not expected that the calibration of LISA will be terrible, but uncertainties should be much smaller than unity. Typical uncertainties in the astrophysical rates of LISA events are one or more orders of magnitude, which will completely dominate over any impact of calibration errors. Finally, the measured SNR also depends on the assumed PSD, $S_h(f)$, and so will also be significantly affected by noise modelling uncertainties, which are likely to be much larger. It therefore does not make much sense to use fluctuations in the measured SNR to place a requirement on calibration if the noise model is fixed. A more robust way to place requirements on calibration is to limit their impact on parameter estimation for detected sources. We will describe how this can be done in the next section.

\subsection{Impact of calibration error on parameter estimation}\label{sec_PE}
%\subsubsection{Modified Fisher Matrix} 
The impact of calibration errors on parameter estimation can be assessed in two different ways. One approach is to assume that the analysis of the data is based on taking $C(f) \equiv 1$, while the true data generating process is described by Eq.~(\ref{eq:calmod}). When the model used in data analysis differs from reality you expect to obtain biases. Calibration error requirements can be set by estimating the induced biases and requiring them to be smaller than uncertainties arising from instrumental noise fluctuations. A second approach is to include additional calibration parameters as part of the model used to fit the data. In that case, provided the model is sufficiently flexible, we would not expect to see biases in parameter estimates, but we would expect that parameter posteriors would be broadened to reflect the additional degrees of freedom coming from the calibration model. We will take the second approach in this paper, but we describe the first approach in Appendix~\ref{sec_bias}, and show how it coincides with the second approach when calibration errors are small.

To implement the second approach, we modify the log-likelihood used in parameter estimation so that it includes the calibration error
\begin{equation}
\log {\cal L} = -\frac{1}{2} \int_0^\infty \frac{|\tilde{d}(f) - C(f|\vec\mu) \tilde{h}(f|\vec\theta)|^2}{S_n(f)} \, {\rm d}f, \label{eq:modlike}
\end{equation}
where $\vec\theta$ are the parameters of the gravitational wave source, as before, and $\vec\mu$ are the parameters describing the calibration error model, in this case the parameters of the amplitude and phase calibration splines. 
We assume that the noise spectral density is known and this allows us to drop constant terms. If we wanted to model the noise as well we would need to include those terms. When analysing observed LISA data using Bayesian inference methods, the calibration error model parameters can be sampled at the same time as the parameters describing the source. The resulting posteriors on the source parameters will be marginalised over calibration errors. This procedure is computationally expensive and therefore impractical for setting calibration requirements over a wide range of source types and source parameters. However, we will use this in Appendix \ref{sec_analyticalfisher} to verify the results we obtain using computationally cheaper methods. %\jon{Add forward ref.} \jon{Is this true? Do we include MCMC results? If so, add a forward reference.}

A cheaper way to assess the impact of calibration errors is to use the Fisher matrix formalism. In the absence of calibration uncertainties, the Fisher matrix is defined by
\begin{equation}
\Gamma^\theta_{ij} = \left( \frac{\partial h}{\partial \theta_i} \bigg| \frac{\partial h}{\partial \theta_j} \right).
\label{eq:fisher_def}
\end{equation}
The Fisher matrix, $\Gamma^\theta_{ij}$, provides an estimate of the precision to which the parameters of the source model can be determined from the data. Specifically, the uncertainty in parameter $\theta^i$, $\Delta \theta^i$, can be estimated as $\langle\Delta \theta^i \Delta \theta^j\rangle = (\Gamma^\theta)^{-1}_{ij}$. In the presence of calibration errors, we can now evaluate the Fisher matrix for the modified likelihood given by Eq.~(\ref{eq:modlike}). This Fisher matrix is over the expanded parameter set that includes both $\vec\theta$ and $\vec\mu$. The modified likelihood is related to the standard likelihood through the replacement $\tilde{h}(f) \rightarrow C(f | \vec\mu) \tilde{h}(f|\vec\theta)$. We can thus directly write down the Fisher Matrix for the combined estimation of $\vec\theta$ and $\vec\mu$
\begin{align}
\Gamma &= \left(\begin{array}{cc} \Gamma^{\theta\theta} & \Gamma^{\theta\mu} \\ (\Gamma^{\theta\mu})^T & \Gamma^{\mu\mu} \end{array}\right) \nonumber \\
\mbox{where } \Gamma^{\theta\theta}_{ij} &= \left( C( \vec\mu)\frac{\partial h}{\partial \theta_i} \bigg|  C( \vec\mu) \frac{\partial h}{\partial \theta_j} \right), \nonumber \\
\Gamma^{\theta\mu}_{ij} & = \left(  C( \vec\mu) \frac{\partial h}{\partial \theta_i} \bigg| h(\vec\theta) \frac{\partial C}{\partial \mu_j} \right), \nonumber \\ \Gamma^{\mu\mu}_{ij} &= \left(  h( \vec\theta) \frac{\partial C}{\partial \mu_i} \bigg| h(\vec\theta) \frac{\partial C}{\partial \mu_j} \right).
\end{align}
We anticipate that calibration will be good and therefore that $C(f) \approx 1$. So, we can evaluate this expression for $C(f)=1$ when assessing and setting calibration requirements. 

Knowledge of the accuracy of calibration can be incorporated by imposing a prior on the calibration parameters, $\vec\mu$. When doing numerical marginalisation any prior can be imposed, but in the Fisher matrix formalism it is easiest to work with a Gaussian prior. Using the prior $\vec\mu \sim$N$(0,\Sigma^{\mu\mu})$ (assuming that we centre the parameters such that $\vec\mu = 0$ corresponds to $C(f)=1$), the posterior covariance is given by the inverse of the modified Fisher matrix
\begin{equation}
\Gamma = \left(\begin{array}{cc} \Gamma^{\theta\theta} & \Gamma^{\theta\mu} \\ (\Gamma^{\theta\mu})^T & \Gamma^{\mu\mu} + \Sigma^{\mu\mu} \end{array}\right).
\label{eq:FIM_GWCalib}
\end{equation}
The diagonal elements of the inverse of this matrix provide estimates for the precision with which the corresponding parameters can be measured. The estimated precision of measurement of the waveform parameters accounting for calibration model uncertainty is thus given by the diagonal elements of the matrix
\begin{equation}
%( \Gamma^{\mu\mu} + \Sigma^{\mu\mu} + (\Gamma^{\theta\mu})^T (\Gamma^{\theta\theta})^{-1} \Gamma^{\theta\mu})^{-1}.
(\Gamma^{\theta\theta} - \Gamma^{\theta\mu} (\Gamma^{\mu\mu} + \Sigma^{\mu\mu} )^{-1} (\Gamma^{\theta\mu})^T)^{-1}.
\label{eq:FMinvtheta}
\end{equation}
Given a specification for the calibration model uncertainties, $\Sigma^{\mu\mu}$, these parameter estimation uncertainties can be evaluated and compared to the diagonal elements of $(\Gamma^{\theta\theta})^{-1}$, which are the estimated uncertainties with perfect calibration. If the uncertainties are significantly larger then our lack of knowledge of the calibration is having an impact on our ability to measure the parameters of the system. For the purpose of concreteness we will say that the calibration knowledge, $\Sigma^{\mu\mu}$, is not good enough if the error predicted by Eq.~(\ref{eq:FMinvtheta}) is more than double that predicted by $(\Gamma^{\theta\theta})^{-1}$ for any parameter. Although chosen arbitrarily, this threshold corresponds to the point at which the systematic biases that arise from ignoring calibration uncertainties become comparable to the size of statistical errors, as described in Appendix~\ref{sec_bias}. This is the standard criterion for concluding that a systematic bias is unacceptably large in the context of assessing the impact of waveform modelling errors~\cite{2007PhRvD..76j4018C}.

In section~\ref{sec_SingSourceReq} we will use this formalism to evaluate the impact of calibration uncertainty on parameter estimation for all different types of LISA source. In all cases we will take $\Sigma^{\mu\mu}$ to be a diagonal matrix, so that we assume the calibration uncertainty at each frequency knot is independent of the others, and amplitude and phase uncertainties are also independent. Additionally, we will usually assume that the size of the uncertainties are the same at all knots, so that there is a single parameter $\sigma_C = \sqrt{\Sigma^{\mu\mu}}$ that characterises the size of the amplitude uncertainty and likewise for the phase. This makes the presentation of the results easier, as we can then plot the ratio of the parameter measurement uncertainties with and without calibration uncertainty as a function of the size of the calibration uncertainties $\sigma_C$, making it easier to identify the point at which the error-doubling threshold is reached. We will present results that vary only the amplitude uncertainties or only the phase uncertainties. For all source types, we have also computed results varying both amplitude and phase calibration uncertainties simultaneously. These are not presented explicitly, as in all cases the results were the same. This is because there are certain parameters that are sensitive to amplitude calibration uncertainties and other parameters that are sensitive to phase calibration uncertainties, but none appear to be sensitive to both.
%\jon{I think Etienne also kept the scale of amplitude and phase calibration the same, but we allowed these to vary separately. We should discuss which results we want in the paper, and whether additional calculations/plots are needed. We also have some results in which we varied the uncertainty at one knot only. I think these could be worth including as they illustrate some useful points. If we include them we should add a forward reference here.}

As a final remark, we note that we can evaluate Eq.~(\ref{eq:FMinvtheta}) with $\Sigma^{\mu\mu} \rightarrow 0$, and obtain finite  uncertainties for many parameters. This corresponds to the limit in which we assume that the calibration is completely unknown and that our data analysis procedure attempts to estimate the calibration simultaneously with the source parameters. The fact that source parameters can still be estimated is because the assumed calibration uncertainty does not look very much like a gravitational waveform and hence can be distinguished from a gravitational wave source. This will be discussed further in section~\ref{sec_GWcalibrators}.

%It is straightforward to evaluate this for other choices, which could provide a useful cross-check, although in the regime where calibration is not limiting measurement precision we do not expect a significant difference. Under this assumption the matrix $\Gamma^{\theta\theta}_{ij}$ is the same as in the case where calibration errors are ignored, and hence is what (future) Figures of Merit codes will be computing. The other matrices can be computed at the same time and this is straightforward if we are using a calibration model that is differentiable, like the spline model proposed here.

\section{Calibration requirements for individual sources}
\label{sec_SingSourceReq}
In this section we will use the formalism described in the preceding section to assess the calibration requirement that arises from requiring that we are able to do accurate parameter estimation on individual resolvable sources. We will consider each of the main categories of resolvable source expected to be observed by LISA.

\subsection{Galactic binaries}
\subsubsection{Definition}
Among the gravitational wave sources expected to be observed by the LISA mission, the most common will come from galactic binaries (GB). Binaries of compact stellar remnants, typically white dwarfs but also neutron stars and black holes, with orbital periods of around 1 hour in the Milky Way emit gravitational waves at millihertz frequencies that LISA will be able to observe. Several tens of millions of these binaries are expected to be present in the LISA data. The majority of these will form an unresolvable astrophysical foreground between $0.5$ mHz and $3$ mHz, but we still expect to individually resolve as many as several tens of thousands of sources.  Various approaches have been proposed to identify and characterise these sources \cite{Crowder:2006eu, Blaut:2009si, Littenberg:2011zg, Littenberg:2020bxy}. As for all sources, we will assess the impact of calibration errors on parameter estimation for individual galactic binaries. In principle, calibration errors could also lead to the incomplete subtraction of binaries from the data and hence lead to an increase of the galactic confusion noise, with a knock-on impact on the characterisation of other LISA sources. We will not consider this here, as we expect the effect to be small in the regime where individual source parameter estimation is unaffected by calibration error.

To represent gravitational waves from galactic binaries we will use the model provided by the LISA Data Challenge \url{https://lisa-ldc.lal.in2p3.fr/}, including all three TDI data channels, $A,E,T$. %\st{These combinations take into account the response of the instrument.} 
The GW signal from a GB is characterized by 8 parameters. The ecliptic latitude $\lambda$, the ecliptic longitude $\beta$, the inclination $\iota$, the polarization $\psi$ and the amplitude $\mathcal{A}$ are extrinsic parameters that describe the position and orientation of the source with respect to LISA. The remaining three parameters are intrinsic and determine the temporal evolution of the plus $h_{+}$ and cross $h_{x}$ polarizations of the GW as follows:
\begin{equation}
\begin{aligned}
    h_{+}^S &= \mathcal{A} \left(1+\cos^2\iota\right) \cos(\Phi(t)), \nonumber \\
    h_{x}^S &= 2 \mathcal{A} \cos\iota \sin(\Phi(t)), \nonumber \\
    \Phi(t) &= 2\pi f_0 t + \pi \dot{f} t^2 - \phi_0 
\end{aligned}
\label{eq:GB_WF}
\end{equation}
where $f_0$ is the GW frequency given at an initial moment $t_0$, $\dot{f}_0$ is the frequency first order derivative and $\phi_0$ is the initial GW phase.  The amplitude depends on the binary chirp mass $\mathcal{M}_c$, the frequency $f_0$ and the distance $D$ :
\begin{equation}
    \mathcal{A} = \frac{\left(G \mathcal{M}_c\right)^{\frac{5}{3}}\left(\pi f_0\right)^{\frac{2}{3}} }{c^4 D} .
\label{eq:GB_Amp}
\end{equation}
The chirp mass is related to the masses of the individual components of the binary, $m_1$ and $m_2$, via $\mathcal{M}_c = m_1^{\frac{3}{5}} m_2^{\frac{3}{5}}/(m_1+m_2)^{\frac{1}{5}}$. For a binary evolving purely under GW emission, the frequency derivative and chirp mass are related through
\begin{equation}
\dot{f}_0= -\frac{96}{5} \frac{\pi^{\frac{8}{3}}}{c^5} \left(G \mathcal{M}_c \right)^{\frac{5}{3}} f_0^{\frac{11}{3}}
\end{equation}
but there are other processes, such as mass-transfer, that can drive frequency evolution in galactic binaries, so we do not use this relation to interpret $\mathcal{A}$ in terms of $D$.

For typical galactic binaries, $\dot{f}_0 \lesssim 10^{-15}$, so the GW signal from a GB is almost monochromatic with a slow drift in the frequency over years of observation, although it is modulated (time-varying Doppler shift) due to LISA's orbital motion.  %This property makes GBs an ideal source to estimate the impact of the calibration error function on the different parameters of the gravitational wave. 
Some galactic binaries, referred to as verification galactic binaries (VGB), have distances that have been well measured using electromagnetic observations. These systems could be used to directly measure the amplitude calibration error and its evolution with time. This will be discussed further in section~\ref{ref_GB_calibrator}.

\subsubsection{Calibration uncertainty Fisher matrix for galactic binaries}
In order to understand which of the galactic binary parameters is most likely to be affected by calibration uncertainty, we first look at the structure of the Fisher matrix for the simultaneous estimation of the calibration uncertainty and GB parameters, computed using the formalism described in section \ref{sec_PE}. We consider a single choice for the GB parameters, but the structure of the Fisher matrix does not depend on this choice. 
%In this section, we will study the effect of the calibration error function on the waveform parameters of a galactic binary using the Fisher matrix . Although the method implemented for this study works for any GB, we choose to present the result for a single galactic binary for brevity's sake. 

We consider a GB with the same parameters as the first VGB (called AM-CVn) present in the LDC catalog (\url{https://gitlab.in2p3.fr/korol/lisa-verification-binaries} and references therein). The parameters are $f_0=1.9\times10^{-3}$, $\dot{f}_0 = 6.1 \times 10^{-18}$, $\mathcal{A}=2.8 \times 10^{-22}$, $\iota=0.8$, $\lambda=0.7$, $\beta=3.0$, $\psi=2.2$ and $\phi_0=6.1$. The waveform is generated for a 4 year duration with LISA analytical orbits given in the LDC manual (\url{https://lisa-ldc.lal.in2p3.fr/static/data/pdf/LDC-manual-001.pdf}), noise according to the first version of the Scientific Requirement Document (SciRDv1) \url{https://www.cosmos.esa.int/documents/678316/1700384/SciRD.pdf} and with instrumental response computed using second order time delay interferometry (TDI2.0).

\begin{figure*}[h!]
    \begin{centering}
    \includegraphics[width=\textwidth]{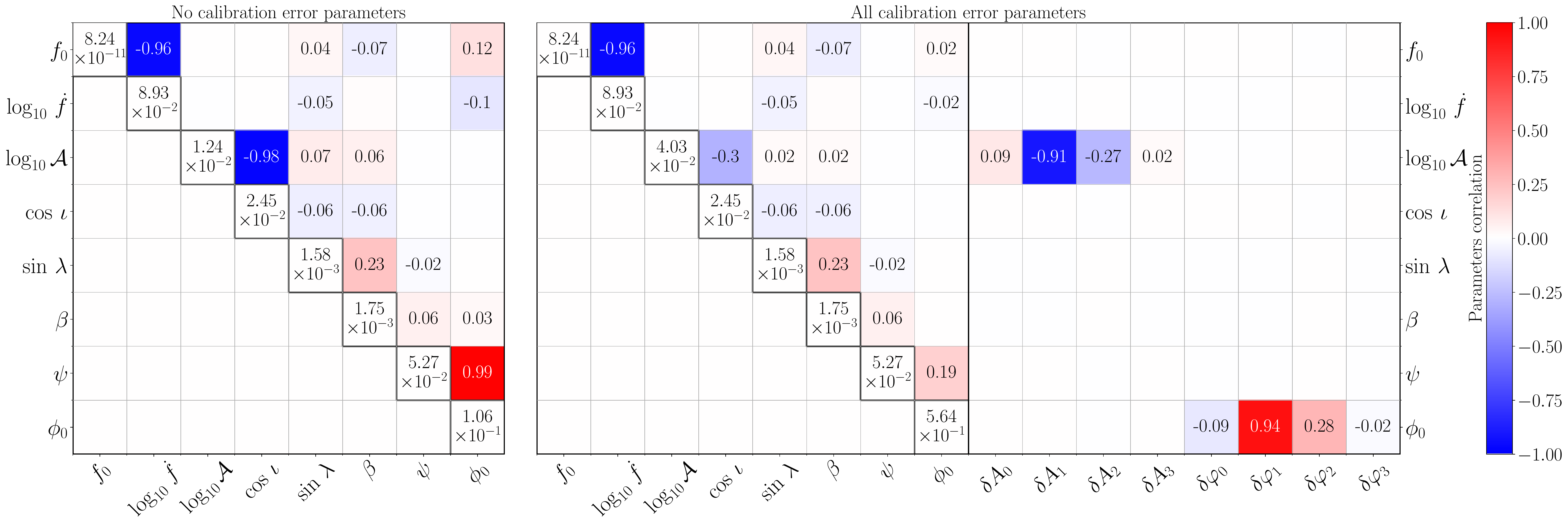}
    \end{centering}
    \caption{Inverse of the Fisher matrix for a single GB with (right) and without (left) a calibration error function. The diagonal elements provide uncertainties in the corresponding parameters, while the off-diagonal elements indicate the degree of correlation in the estimates of pairs of parameters. As we are primarily interested in the impact of calibration on the determination of the source parameters, for clarity we do not show the calibration-calibration sub-matrix. Off-diagonal elements that are smaller than $0.01$ are left blank.}
    \label{fig:GB_FIM}
\end{figure*}

As discussed in section \ref{sec_calibrationmodel}, the full calibration error model we use here consists of two cubic spline functions, defined by four amplitude error parameters $\{\delta A_0,\delta A_1,\delta A_2,\delta A_3\}$ and four phase error parameters $\{\delta \varphi_0,\delta \varphi_1,\delta \varphi_2,\delta \varphi_3\}$. The cubic spline is constructed as a function of the logarithm of frequency to ensure smoothness over the full LISA frequency range. %\jon{Revisit if necessary after addressing cubic spline versus cubic question above.}
%\jon{I think some stuff was cut out of appendix B. It is true that the above expression holds between any two knots, and hence is valid for GBs which basically remain at a fixed frequency and don't cross knots. But appendix B doesn't give the relationship to the knot weights. We can re-instate this there, or add this in section IIA. TO DISCUSS.} 
The elements of the inverse Fisher matrix, with and without the inclusion of calibration errors, are given in figure \ref{fig:GB_FIM}. The results with calibration errors take the calibration error uncertainty to be $\sigma_C=0.1$. The diagonal elements of the inverse Fisher matrix provide an estimate of the precision with which the corresponding parameter can be determined, while the off-diagonal elements indicate the amount of correlation in the uncertainties between pairs of parameters.

Comparing the diagonal elements in the left and right panels we see that the only two parameters for which the uncertainties change significantly when calibration uncertainties are included are $\mathcal{A}$ and $\phi_0$. This conclusion is supported by the off-diagonal elements which show that these two parameters are the only ones that are highly correlated with the calibration model parameters, specifically $\delta\mathcal{A}_1$ and $\delta \varphi_1$. This makes sense physically, since a change in $\mathcal{A}$ represents a constant change in the amplitude of the signal, and $\delta\mathcal{A}_1$ does the same thing, at the frequency of this source. Similarly a change in $\phi_0$ is a constant change in phase, which can also be accomplished by a change in $\delta\varphi_1$.

%Without the calibration error function (left of figure \ref{fig:GB_FIM}), all the correlations (and anti-correlations) represented by colored tile are expected for such GW source (frequency/frequency derivative, amplitude/inclination, initial phase/polarization). Once the calibration error is taken into account (right of figure \ref{fig:GB_FIM}), the amplitude error parameters are correlated with the waveform amplitude while the phase error parameters are correlated with the waveform initial phase. 

\subsubsection{Setting a calibration requirement for galactic binary parameter estimation}
\begin{figure*}[h!]
    \centering
    \includegraphics[width=\textwidth]{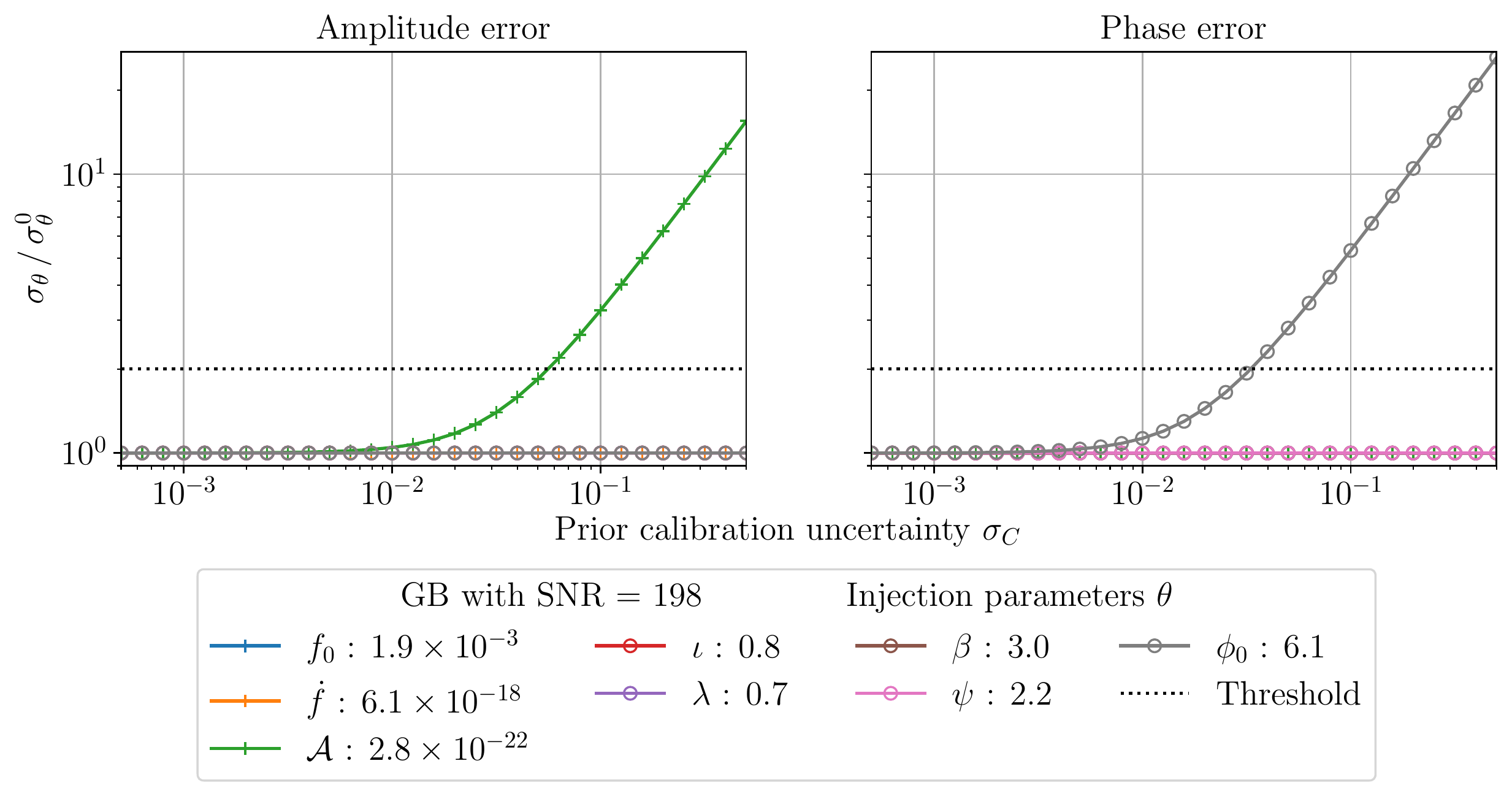}
    \caption{Ratio between the uncertainty of a waveform parameter with ($\sigma_{\theta}$) and without (${\sigma_{\theta}}_0$) the calibration error function as a function of the size of the calibration error uncertainty $\sigma_C$. %The uncertainty of the parameters in the absence of the calibration error function $\sigma_0$ is given in the legend. \jon{Not any more....} 
    The red line corresponds to a doubling of the parameter uncertainty which is our threshold for the calibration uncertainty being unacceptably large. In the left hand panel we vary only the amplitude calibration uncertainty, and in the right panel we vary only the phase calibration uncertainty. %The linear increase of the uncertainty error induces a linear increase of the uncertainty of the phase or/and the amplitude coming respectively from the phase error or/and the amplitude error.
    } 
    \label{fig:GB_uncertainty}
\end{figure*}

In the previous section we saw the precision of parameter estimation changed for a particular choice of calibration uncertainty. To set a calibration requirement, we must vary the size of the calibration uncertainty to see how large this can be before parameter estimates become significantly degraded. These results are shown in Figure~\ref{fig:GB_uncertainty} when varying only the amplitude calibration uncertainty or only the phase calibration uncertainty. Results varying both simultaneously are the same and are therefore not shown. 
%In order to estimate the impact of the calibration error on the wave parameters, we need to project the Fisher matrix before inverting it. By varying the value of the uncertainty of the calibration error, it is then possible to estimate the change in a waveform parameter uncertainty that would be correlated with some of the calibration error parameters.
Consistent with the form of the inverse Fisher matrix in Figure \ref{fig:GB_FIM}, we find that the only parameter impacted by phase calibration uncertainty is the initial phase, $\phi_0$, and the only parameter affected by amplitude calibration uncertainty is the amplitude/distance. In each case, the precision of parameter estimation begins to degrade when the size of the calibration uncertainty is comparable to the precision with which the affected parameter can be measured in the absence of these uncertainties. Requiring that the parameter uncertainty no more than doubles sets a requirement that the amplitude calibration uncertainty is no bigger than a few percent and the phase calibration uncertainty is no bigger than $3\times10^{-2}$.

Once the uncertainty starts to increase it varies roughly linearly with the size of the calibration uncertainty. This behaviour can be understood analytically, which is discussed further in appendix~\ref{sec_analyticalfisher}. While these results were based on the Fisher matrix, they can also be verified by doing full Bayesian parameter estimation with MCMC methods. We have done this for the case of galactic binaries and the MCMC results verify the conclusions of the Fisher matrix. See Figure~\ref{fig:GB_ana} in appendix~\ref{sec_analyticalfisher}.

The stated calibration requirements are valid for this particular choice of GB parameters, but we would find a different requirement if we considered different parameter choices. This dependence on the source parameters can be understood from the analytic results described in appendix~\ref{sec_analyticalfisher}, but in general we expect the impact of calibration uncertainties to be greatest on sources for which we can determine the parameters most precisely. These are typically the highest signal-to-noise ratio systems. We expect the system considered here to lead to one of the most stringent constraints, so we would not expect to need more than a factor of a few more stringent calibration requirement in the worst case.

%It remains important to remember that the uncertainty in the absence of error is roughly inversely proportional to the ratio of the source considered, making this threshold "source-dependent". Another way of defining a constraint on the calibration error for GBs can be found in the linear coefficient $K_{\theta}$ represented \stas{wasn't it called "s" before?} in the legend of figure \ref{fig:GB_uncertainty} which depends essentially on the center frequency of the GB considered. With such an evaluation tool, it is therefore possible to use this simple linear law to define a calibration error agnostic threshold.

\subsection{Massive black hole binaries}
\subsubsection{Definition}
The mergers of massive black hole binaries (MBHBs) are among the loudest gravitational wave sources expected to be observed by the LISA mission. Most galaxies contain massive black holes at their centres and MBHBs are expected to form following the merger of their host galaxies. Such systems are believed to form even very early in the history of the Universe, and LISA can observe systems of suitable mass ($\sim 10^4M_\odot$--$10^7M_\odot$) to very high redshift ($z\sim 20$). LISA is expected to observe several tens of sources and these observations will enable us to better understand the formation processes of these binary systems, their environments and to discriminate between different scenarios of seed MBH formation.

In this analysis, we model the source in the Fourier-domain using the IMRPhenomD model \cite{Khan:2015jqa}, which represents coalescing black holes in which the spins are parallel to the orbital angular momentum. We construct the LISA TDI response in the same way as for the galactic binaries. The GW signal from a MBHB is characterized by 11 parameters. There are the same five extrinsic parameters that were described for the GB model: the ecliptic latitude $\beta$, the ecliptic longitude $\lambda$, the inclination $\iota$, the polarization $\psi$ and the distance $D$. % \etienne{Add units} are extrinsic parameters that allow to position the source with respect to LISA. 4 of the 
The additional intrinsic parameters are the chirp mass, $\mathcal{M}_c$, the symmetric mass ratio, $\eta = m_1 m_2/(m_1+m_2)^2$, the spins of the two components, $\chi_{1,2}$, the phase at coalescence, $\phi$, and the coalescence time, $\tau$. The bandwidth of an MBHB signal in the frequency domain is broad, in contrast to the narrow bandwidth of a GB, which we expect to help us distinguish MBHB signals from calibration errors.

\subsubsection{Calibration uncertainty Fisher matrix for massive black hole binaries}
\begin{figure*}[h!]
    \centering
    \includegraphics[width=\textwidth]{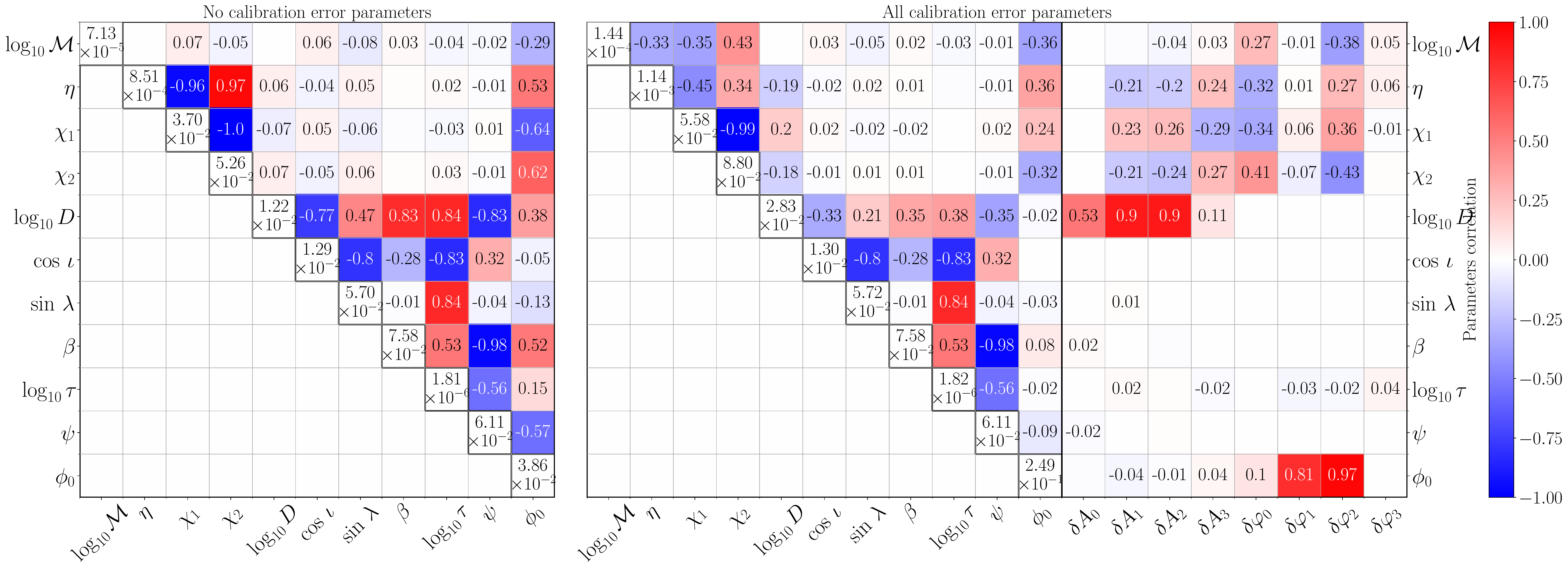}
    \caption{As Figure~\ref{fig:GB_FIM} but now for a MBHB source.
    }
    \label{fig:MBHB_FIM}
\end{figure*}

As for the GB case, we first look at the correlations in the Fisher matrix for simultaneous estimation of calibration errors and MBHB parameters. The inverse Fisher matrix is shown in Figure \ref{fig:MBHB_FIM} for a typical MBHB source with parameters $\mathcal{M}_c=7.82\times 10^5M_\odot$, $\eta=0.2$, $\chi_1=\chi_2=1.0$, $D=1.42 \times 10^5$Gpc, $\iota=2.3$, $\lambda=-0.6$, $\beta=0.6$, $\tau=4.8 \times 10^6$s, $\psi=3.1$ and $\phi_0=4.3$. We again take the prior on the calibration uncertainty to have a width $\sigma_C=0.1$. 

Comparing the diagonal elements in the left and right hand figures we see that the inclusion of calibration uncertainties has a moderate impact on the measurements of several of the intrinsic parameters -- $\mathcal{M}_c$, $\eta$, $\chi_1$, $\chi_2$ -- and a more significant impact on the measurement of $\phi_0$ and $D$. This is to be expected from looking at the off-diagonal elements, which show that the measurement of these four intrinsic parameters is mildly ($\sim 0.3$) correlated with the measurement of the calibration uncertainty parameters, while the measurements of $\phi_0$ and $D$ are strongly ($\sim 0.9$) correlated with the phase and amplitude calibration uncertainty parameters respectively. These results make sense -- changes in the amplitude and $\phi_0$ impact the MBHB signal in the same way that they impact the GB signal and so we expect the same degeneracy with the calibration parameters that we saw in the GB case. The other intrinsic parameters primarily affect the phase evolution of the MBHB signal and so we might expect a mild degeneracy with the phase calibration uncertainty. 
%Once the calibration error is taken into account (right of figure \ref{fig:MBHB_FIM}), the correlations between calibration error parameters and waveform parameters are not as trivial as the GB case. The detailed study of the correlations would bring little to the discussion, nevertheless it is possible to notice the obvious correlation between distance and amplitude error. Since the amplitude and phase of an MBHB depend on the intrinsic parameters, we expect their coupling with both the amplitude and phase errors. 

\subsubsection{Setting a calibration requirement for MBHB parameter estimation}
\begin{figure*}[h!]
    \centering
    \includegraphics[width=\textwidth]{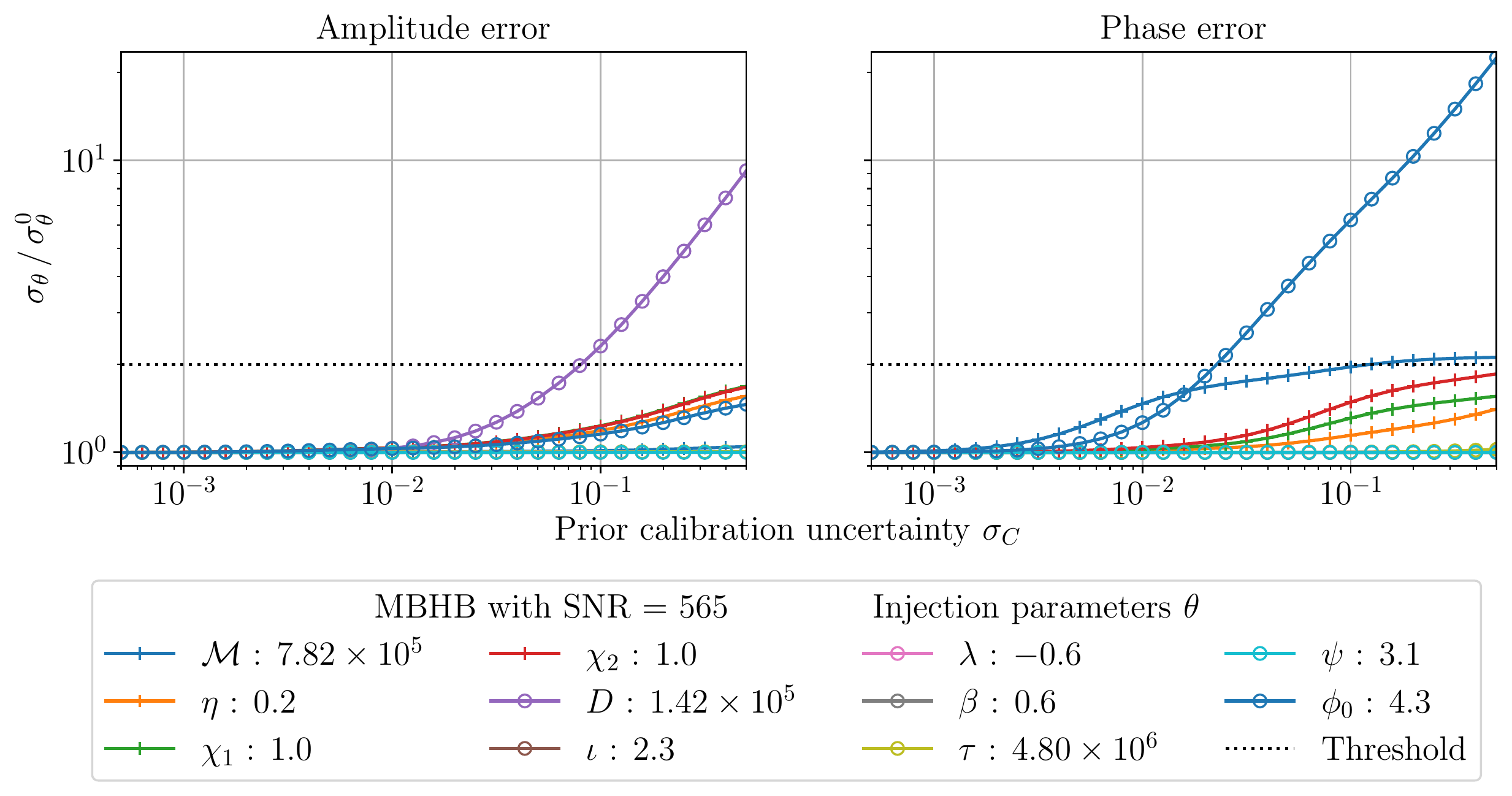}
    \caption{As Figure~\ref{fig:GB_uncertainty}, but now for the MBHB source shown in Figure~\ref{fig:MBHB_FIM}.
    %Ratio between the uncertainty of a waveform parameter with ($\sigma_{\theta}$) and without (${\sigma_{\theta}}_0$) the calibration error function as a function of the calibration error uncertainty $\sigma_C$. The uncertainty of the parameters in the absence of the calibration error function $\sigma_0$ is given in the legend. The red line corresponds to a two times increase of the waveform parameter uncertainty that could be detrimental for some parameters. From left to right, the calibration error function incorporates only the phase error, only the amplitude error or both errors.
    }
    \label{fig:MBHB_uncertainty}
\end{figure*}
We now explore the variation in the parameter measurement uncertainty for MBHBs as we vary the size of the calibration uncertainty. These results are shown in Figure~\ref{fig:MBHB_uncertainty}. As expected we see that once the amplitude calibration uncertainty is comparable to the precision with which the distance can be measured, it starts to limit our ability to measure distance. Once the phase calibration uncertainty is sufficiently large, it has a similar impact on measurements of the phase offset, $\phi_0$. In this case we also see a mild impact on the other intrinsic parameters -- $\mathcal{M}_c$, $\eta$, $\chi_1$ and $\chi_2$. Once again, our ability to measure these parameters starts to degrade once the calibration uncertainties become comparable with the expected precision on these parameters in the absence of calibration uncertainty. For these parameters, however, the uncertainty does not continue to increase, but levels off at about twice the uncertainty in the absence of calibration errors. This happens because the degeneracy is not perfect in this case -- the phase evolution of an MBHB waveform does not look like a calibration uncertainty of the form modelled here and so the two effects can be distinguished even with no prior on the size of the calibration error, with only mild confusion.

For this source we deduce a calibration requirement of $\sim 8\%$ in amplitude and $2 \times 10^{-2}$ in phase to ensure the measurement uncertainties are no more than doubled. Once again, this conclusion will be source dependent but this source is relatively high signal-to-noise ratio and so we would not expect the results to change by more than a factor of a few for other sources.

%\jon{Add comments here on the fact that this source is unusually good compared to a typical source drawn from the population. Also comment on weak lensing and the fact that the calibration requirement need not be any more stringent than the weak lensing limit.}
%\etienne{I changed the source to be representative of what is expected.}
%TO HERE

\subsection{Stellar-origin black hole binaries}
\subsubsection{Definition}
Stellar-origin black hole binaries (SBHBs) refer to binaries of compact objects formed as the end point of binary stellar evolution. These are the systems from which ground-based detectors are observing gravitational waves emitted during the final stages of inspiral and merger. For binaries with sufficiently large component masses ($\sim 30M_\odot$), LISA will observe gravitational waves emitted by these systems as the gradually inspiral, $\sim10$ years before merger. The number of these events that will be observed is somewhat uncertain, as it depends critically on the high-mass end of the SBHB mass spectrum, which is not yet well constrained. However, they are of great scientific interest because of the prospect of multi-band observations of the same system with both LISA and ground-based instruments.

In this work we use the same model for gravitational waves from SBHBs as we do for MBHBs, the aligned-spin IMRPhenomD \cite{Khan:2015jqa} model, as implemented in the LISA Data Challenge. The parameters characterising these systems are the same as those for the MBHBs, the only difference is that the chirp mass, $\mathcal{M}_c$, and distance, $D$, take different ranges. As for MBHBs, these systems are relatively broad in the frequency domain which should help to distinguish these systems from calibration errors. Unlike MBHBs, LISA is expected to observe SBHBs with relatively low signal-to-noise ratio and measure the parameters with relatively poorer precision. This should also limit the impact of calibration errors.

%The GW signal from SBHB is characterized by 11 parameters. The ecliptic latitude $\beta$, the ecliptic longitude $\lambda$, the inclination $\iota$ and the polarization $\psi$ and the distance $D$ are extrinsic parameters that allow to position the source with respect to LISA. 4 of the remaining parameters are intrinsic with the chirp mass $\mathcal{M}_c$ and symmetric mass ratio $\eta$ and two spins $\chi_{1,2}$. Finally, the initial phase $\varphi$ and the initial frequency $f_0$ allows to describe the waveform with respect to the observer.

\subsubsection{Calibration uncertainty Fisher matrix for stellar-origin black hole binaries}
\begin{figure*}[h!]
    \centering
    \includegraphics[width=\textwidth]{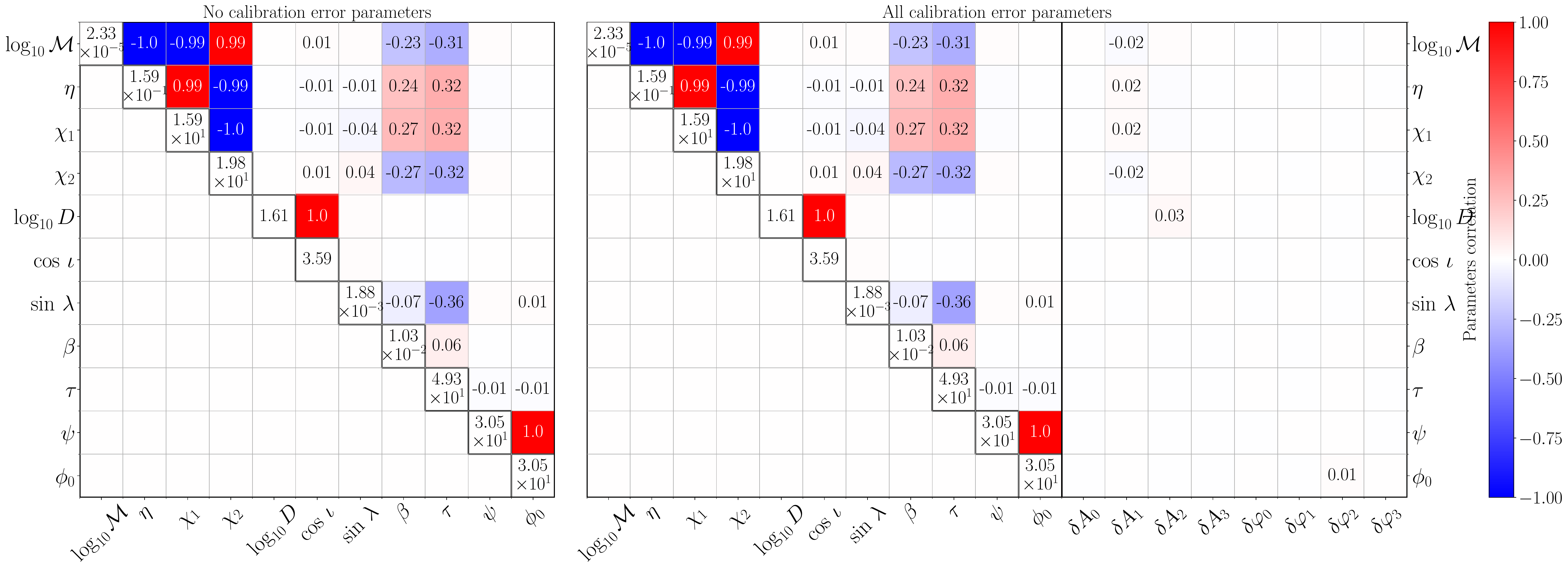}
    \caption{As Figure~\ref{fig:GB_FIM} but now for a SBHB source.
    %Inverse of the Fisher matrix for a single SBHB with (right) and without (left) a calibration error function. The diagonal contains the uncertainty of the different parameters, while the off diagonal parameters represents the parameters correlation. Being null, the correlation between the calibration error parameters are not shown in the right matrix.
    }
    \label{fig:SBHB_FIM}
\end{figure*}
In Figure~\ref{fig:SBHB_FIM} we again show the elements of the inverse Fisher matrix with and without the inclusion of the calibration uncertainty parameters. These are shown for a SBHB source with parameters $\mathcal{M}_c=47.9 M_\odot$, $\eta=0.2$, $\chi_1=\chi_2=0.0$, $D=742.82$Mpc, $\iota=0.4$, $\lambda=2.0$, $\beta=1.2$, $\tau=2.1 \times 10^8$s, $\psi=-3.3$ and $\phi_0=2.7$. Once again we take $\sigma_C=0.1$.  Comparing the diagonal elements of the Fisher matrix in the left and right hand panels we find in this case that there is essentially no change in the precision of parameter estimation when simultaneously fitting for the calibration model. This is supported by the off-diagonal elements of the Fisher matrix which show that there is almost no correlation between the parameters of the signal model and  those of the calibration error model.
%Once the calibration error is taken into account (right of figure \ref{fig:MBHB_FIM}), the correlations between calibration error parameters and waveform parameters are easier to predict than in the case of an MBHB. In the LISA frequency band, the SBHB sources are only in the inspiral phase of the GW which limits the effect of an error of phase on the intrinsic parameters (see appendix \ref{sec_analyticalfisher}). We find again the obvious correlation between phase error and initial phase as well as error in amplitude and distance.

\subsubsection{Setting a calibration requirement for SBHB parameter estimation}
\begin{figure*}[h!]
    \centering
    \includegraphics[width=\textwidth]{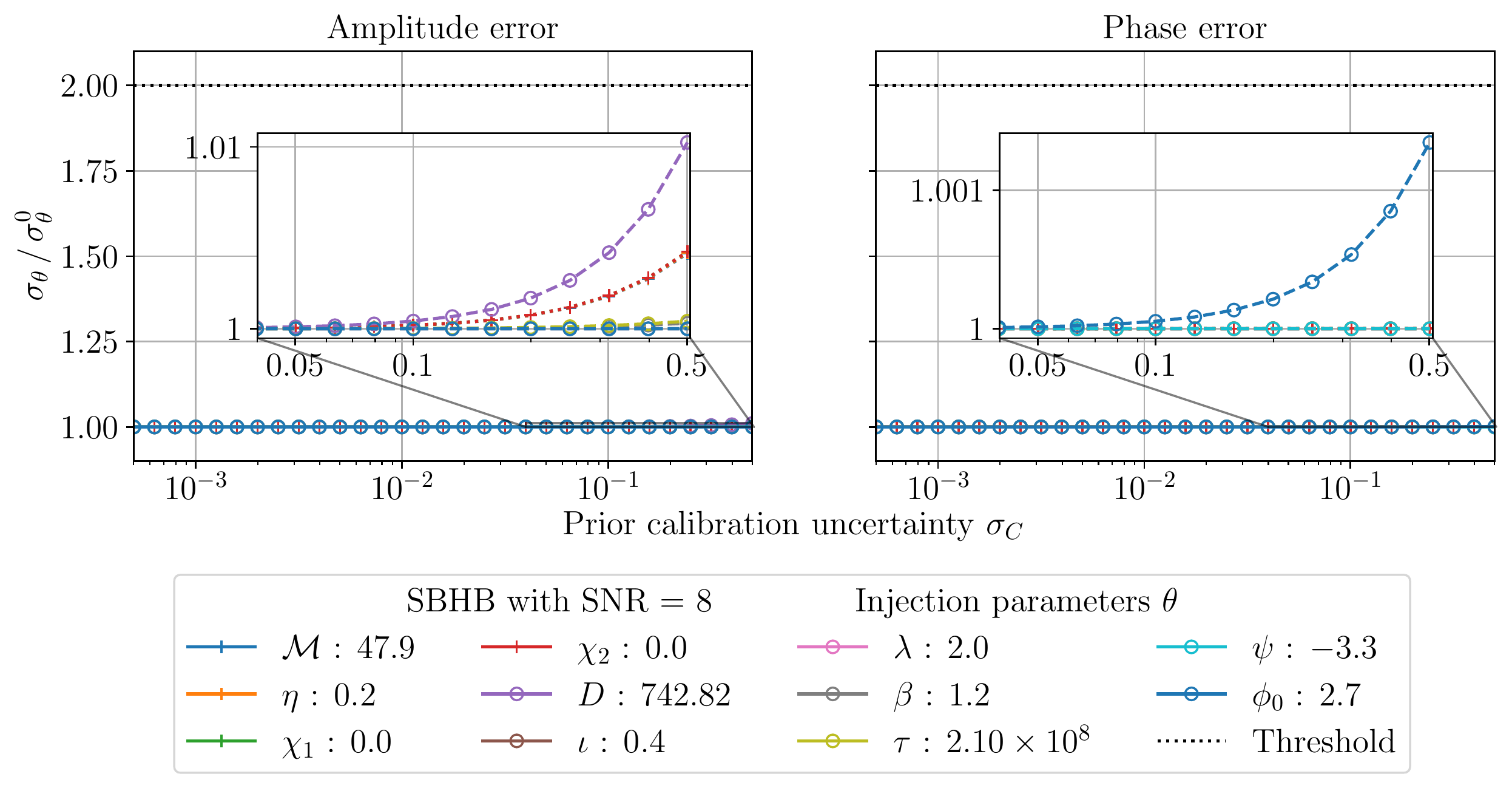}
    \caption{As Figure~\ref{fig:GB_uncertainty} but now for the SBHB source shown in Figure~\ref{fig:SBHB_FIM}.
    %Ratio between the uncertainty of a waveform parameter with ($\sigma_{\theta}$) and without (${\sigma_{\theta}}_0$) the calibration error function as a function of the calibration error uncertainty $\sigma_C$. The uncertainty of the parameters in the absence of the calibration error function $\sigma_0$ is given in the legend. The red line corresponds to a two times increase of the waveform parameter uncertainty that could be detrimental for some parameters. From left to right, the calibration error function incorporates only the phase error, only the amplitude error or both errors. 
    }
    \label{fig:SBHB_uncertainty}
    %\etienne{Useless plot with the cubic spline simulation. Add zoomed-in plots}
\end{figure*}
As before we now vary the size of the calibration uncertainty, $\sigma_C$. The results are shown in Figure~\ref{fig:SBHB_uncertainty}. We see that even for relatively large calibration uncertainties, the parameter measurement precisions remain almost unchanged relative to their values in the absence of calibration uncertainty. The inset panels show that for calibration uncertainties of order unity we begin to see an impact on the measurement of distance and initial phase, but the error increases by less than 1\%. This result is driven by the fact that SBHBs are observed with relatively low signal-to-noise ratio and with relatively poor parameter estimation precision. Artificially increasing the signal-to-noise we observe more of an effect (see appendix~\ref{sec_analyticalfisher}). Based on these results, the characterisation of SBHBs does not appear to impose a requirement on the calibration of the LISA instrument.
% remain under our threshold for any value of the calibration error. This observation is explained by the low signal-to-noise ratio of the SBHBs expected for the LISA mission, which translates into a poor precision on the evaluation of the parameters of the gravitational wave. In order to overcome this limitation, we can artificially reduce the LISA noise thus increasing the SNR revealing the calibration error effect on such sources (see appendix \ref{sec_analyticalfisher}). In this idealized scenario, we are able to semi-analytically confirm the numerical results obtained with the numerical Fisher matrix-based method. 
%For this particular SBHB source and more generally for any type of sources with a low SNR, there would be no detectable effect of the calibration error.

\subsection{Extreme-mass-ratio inspirals}
\subsubsection{Definition}
Extreme-mass-ratio inspirals (EMRIs) are the inspirals of stellar-origin compact objects (COs) into massive black holes (MBHs) in the centres of galaxies. EMRI observations with LISA have great scientific potential, since the compact object typically generates $10^4 - 10^5$ detectable waveform cycles in band, during which time it is orbiting in the strong field region close to the central black hole. The emitted gravitational waves encode a detailed map of the space-time outside the central MBH and offer a unique opportunity to measure the properties, evolution and environment of MBHs 
\citep{barackLISACaptureSources2004,arunMassiveBlackHole2009, barausseCanEnvironmentalEffects2014, gairConstrainingPropertiesBlack2011,gairLISAExtrememassratioInspiral2010,amaro-seoaneIntermediateExtremeMassRatio2007}, to test for deviations from General Relativity (GR)~\citep{gairTestingGeneralRelativity2013,barackUsingLISAEMRI2007} and to constrain cosmological parameters \citep{macleodPrecisionHubbleConstant2008,laghi2021gravitational}.
Event rates are highly uncertain, but LISA could observe between a few and a few hundred EMRIs over the mission duration \citep{gairEventRateEstimates2004,gairProbingBlackHoles2009,babakScienceSpacebasedInterferometer2017,pan2021formation,amaro-seoaneIntermediateExtremeMassRatio2007, amaro-seoaneRelativisticDynamicsExtreme2018}.

The great scientific potential of these astrophysical sources relies on being able to make very precise measurements of their parameters and this could therefore be significantly impacted by calibration uncertainties. To assess this we use two different waveform models: the Analytical Kludge~\citep{barackLISACaptureSources2004} and the Augmented Analytical Kludge as implemented in the Fast EMRI waveform package~\citep{Katz:2021yft,michael_l_katz_2020_4005001,Chua:2015mua,Chua:2017ujo,Fujita:2020zxe,Stein:2019buj}. We use the low-frequency approximation to implement LISA response, as first described in~\cite{cutlerAngularResolutionLISA1998,barackLISACaptureSources2004}. The full EMRI parameter space is seventeen dimensional~\citep{michael_l_katz_2020_4005001}, but for the purposes of this study we restrict attention to seven parameters: the central black hole mass, $M$ [M$_\odot$], the dimensionless spin parameter of the central MBH, $a$, the compact object mass, $\mu$ [M$_\odot$], the initial semi-latus rectum, $p_0$, eccentricity, $e_0$, and cosine of the inclination, $Y_{0}$, of the orbit, and the luminosity distance, $d_L$ [Gpc].

\subsubsection{Calibration uncertainty Fisher matrix for EMRIs}
\begin{figure*}[h!]
    \centering
    \includegraphics[width=\textwidth]{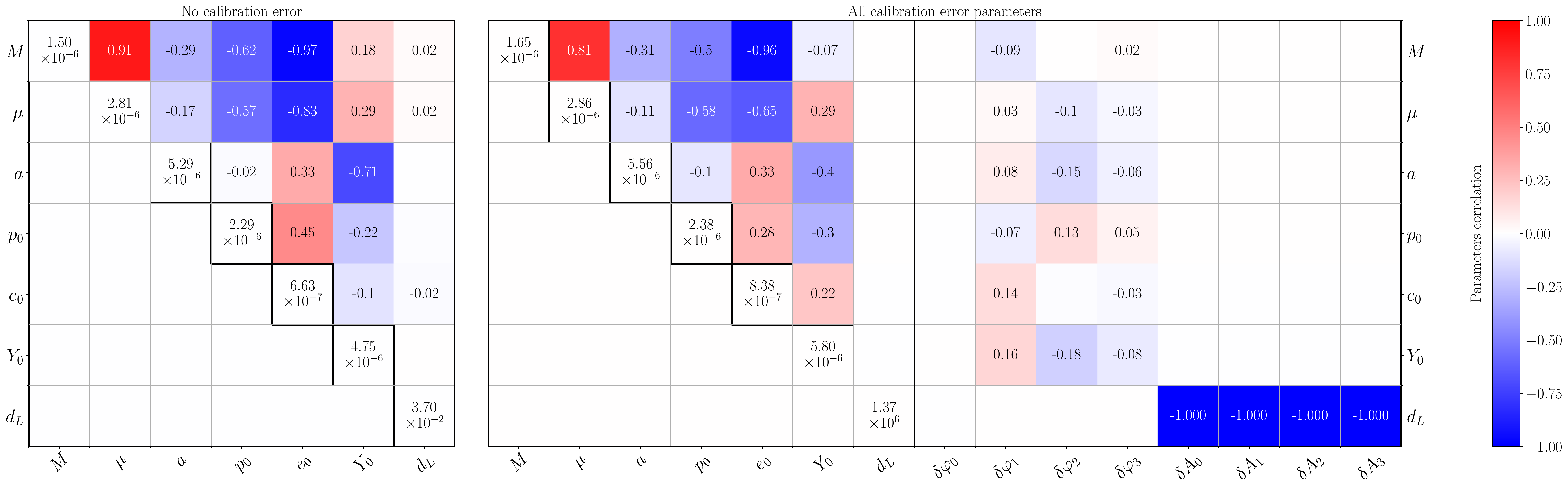}
    \caption{As Figure~\ref{fig:GB_FIM} but now for an extreme-mass-ratio inspiral source.
    %Inverse of the Fisher matrix for a single EMRI with (right) and without (left) a calibration error function. The diagonal contains the uncertainty of the different parameters, while the off diagonal parameters represents the parameters correlation. Being null, the correlation between the calibration error parameters are not shown in the right matrix.
    } %\etienne{Change the caption if needed }\lorenzo{update with the new source of the next figures}
    \label{fig:EMRI_FIM}
\end{figure*}

In Figure~\ref{fig:EMRI_FIM} we show the inverse of the Fisher matrix with and without including calibration uncertainties for an EMRI source with parameters $M=10^6M_\odot$, $a=0.9$, $\mu=23.1 M_\odot$, $p_0=12.0$, $e_0=0.6$, $Y_0=0.7$ and $d_L=1.1$Gpc, observed for 2 years. The EMRI model used for these results was the AAK, but we found the same quantitative results when using the AK model. We again fix $\sigma_C=0.1$. Comparing the diagonal elements of the Fisher matrix on the left and the right we see that distance measurement precision is significantly impacted by calibration uncertainties and there is also a minor impact on measurements of the intrinsic parameters. As in previous cases, this behaviour is reinforced by looking at the off-diagonal elements of the Fisher matrix, which shows weak correlations between measurements of the intrinsic parameters of the source and the phase calibration parameters, and a very strong correlation between measurements of the distance and the amplitude calibration parameters.

\subsubsection{Setting a calibration requirement for EMRI parameter estimation}
\begin{figure*}[h!]
    \centering
        \includegraphics[width=\textwidth]{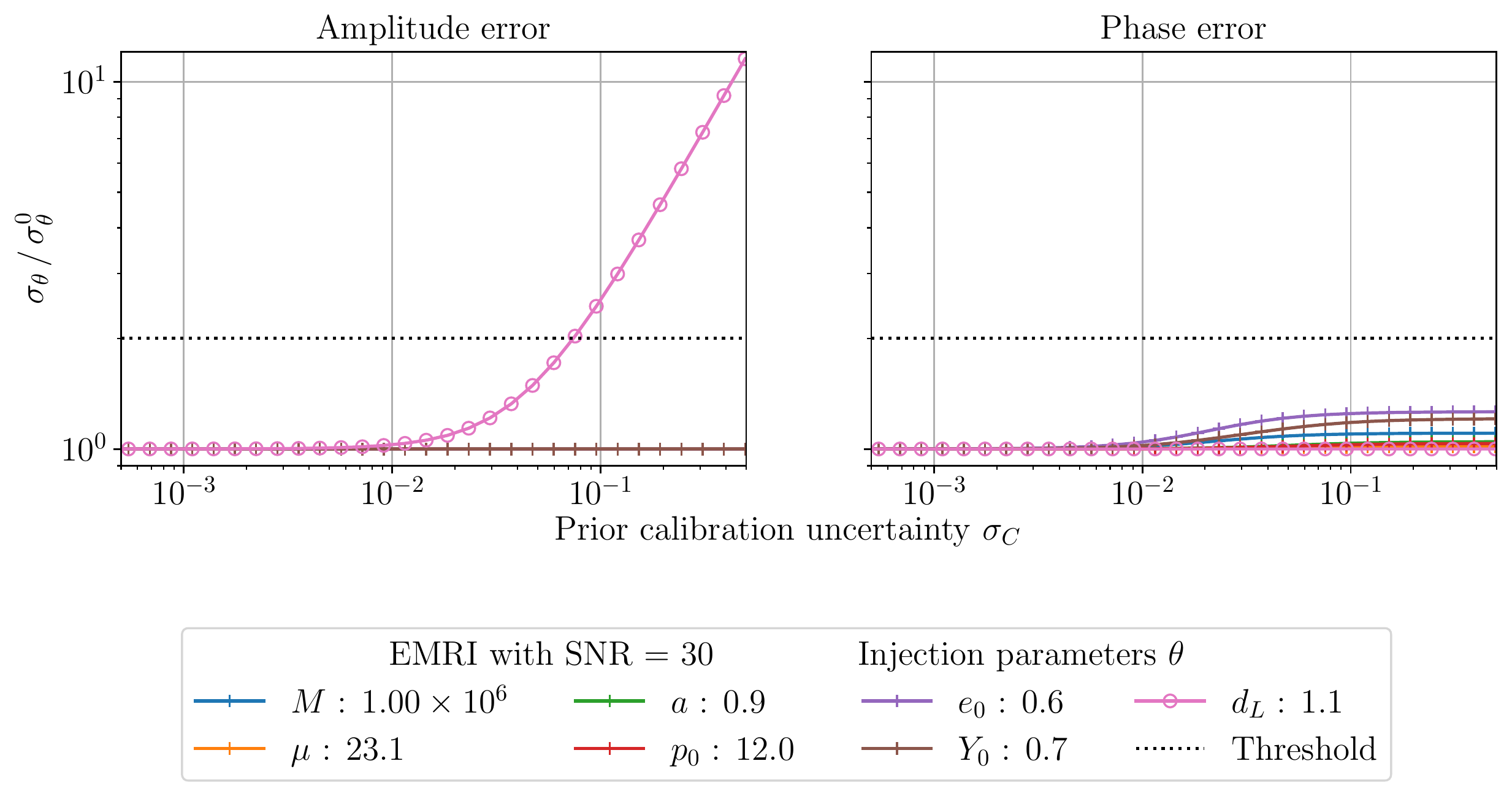}
    \caption{As Figure~\ref{fig:GB_uncertainty} but now for the EMRI source shown in Figure~\ref{fig:EMRI_FIM}. %\jon{Figure legend needs subscripts on $p_0$, $e_0$, $Y_0$.}
    %Ratio between the uncertainty of a waveform parameter with ($\sigma_{\theta}$) and without (${\sigma^0_{\theta}}$) the calibration error function as a function of the calibration error uncertainty $\sigma_C$. The injected values of the source are shown in the legend. The black dotted line corresponds to a two times increase of the waveform parameter uncertainty. In the left plot the calibration error function incorporates only the phase errors, whereas in the right panel only the amplitude errors are included.
    }
     %\etienne{Change the caption if needed, and make the x/y axis label size consistent with previous sources}
    \label{fig:EMRI_calib}
\end{figure*}
We now vary $\sigma_C$, which determines the size of the calibration uncertainties, and determine how the parameter measurement precisions change. These results are shown in Figure~\ref{fig:EMRI_calib} and were again computed using the AAK EMRI model. Results from the AK model show the same quantitative behaviour. These results are consistent with the conclusions from looking at the elements of the inverse Fisher matrix. Once the amplitude calibration uncertainty becomes too large, we see that the distance measurement precision rapidly degrades, ultimately linearly with the size of the calibration uncertainty. The measurement precision of the intrinsic parameters also starts to degrade once the phase calibration uncertainty becomes large enough, but the degradation is limited and the measurement precision is never more than $\sim25\%$ worse than in the absence of calibration uncertainty.

%The prior knowledge on the range of variation of the calibration parameters is encoded in the prior calibration uncertainties $\sigma_C$. As our uncertainty on the calibration errors increases, the precision with which we will be able to recover the parameters degrades. We want to asses when the size of these errors start to inflate the measurement error of the EMRI parameters. Therefore, we estimate the measurement precision of the EMRI and calibration parameters $\theta$ by using the aforementioned Fisher matrix formalism. We denote the parameter measurement error in the presence of calibration errors with $\sigma _{\theta}$ and we compare it to the same error $\sigma^0 _\theta$ calculated in the absence of calibration, i.e. as $\delta C \rightarrow 0$. We show in Figure~(\ref{fig:EMRI_calib}) the impact of calibration errors on the parameter measurement precision of an EMRI system using the AAK waveform model.

These results are consistent with results for other sources and are readily understood. The effect of amplitude calibration error in the observed data is degenerate with the effect of a change in the source distance, so once the uncertainty in the former becomes comparable with the expected measurement precision in the latter it starts to dominate the uncertainty. The intrinsic parameters are measured primarily through their influence on the phase evolution of the source and so we expect them to be somewhat degenerate with phase calibration uncertainty. However, the complex evolution of the phase with  frequency in an EMRI is very unlike the slowly varying phase calibration error we assume here and so the degeneracy is limited, allowing both effects to be simultaneously constrained from the observed data.

\begin{figure*}[h!]
    \centering
    \includegraphics[width=\textwidth]{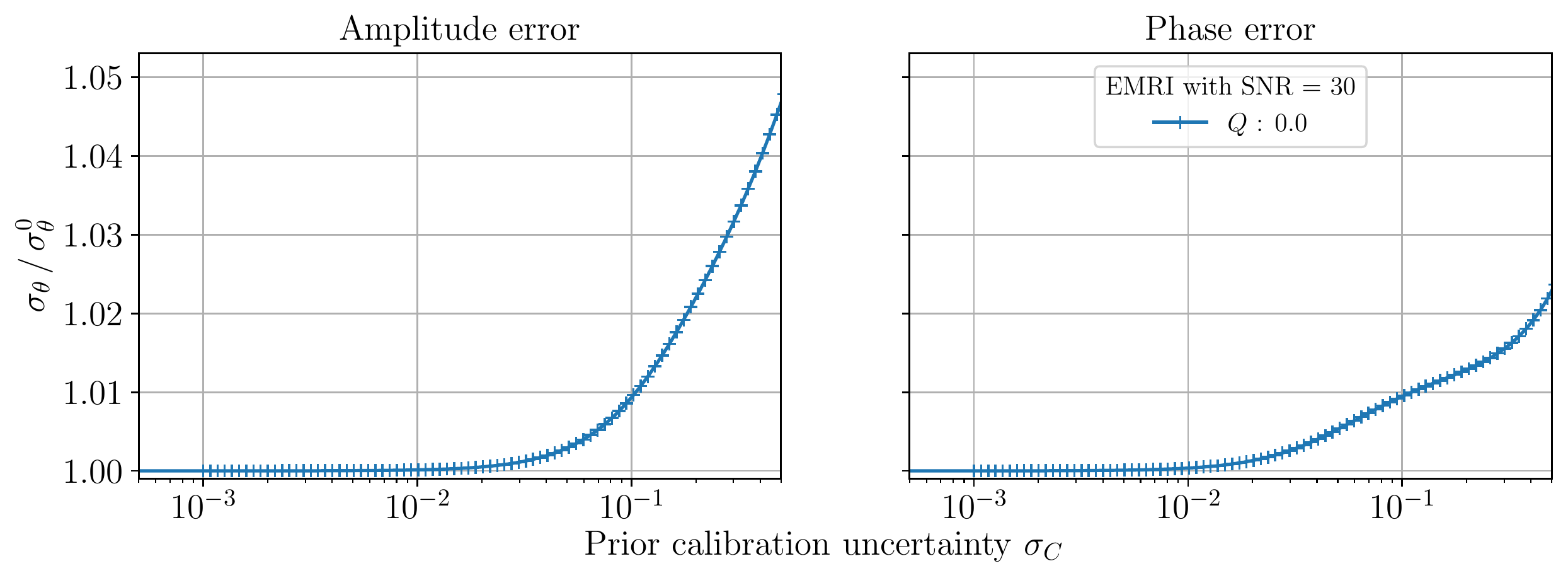}
    \caption{As Figure~\ref{fig:EMRI_calib} but now for the excess quadrupole moment of the extended AK model described in~\citep{barackUsingLISAEMRI2007}. The waveform parameters used here are different to those used for Figure~\ref{fig:EMRI_calib}, but the results are similar for all choices of EMRI parameters.%\jon{Lorenzo to update plot for consistent plotting style.}
    \citep{babakScienceSpacebasedInterferometer2017}
    \citep{emri_Q_2007}
    %\lorenzo{check whether the EMRI parameters are the same of the previous figure or add the values}
    }
    \label{fig:EMRI_excess_quad}
\end{figure*}

One of the primary scientific applications of EMRI observations will be to provide stringent tests that the spacetime structure outside the MBH is consistent with the Kerr metric, as predicted by the no-hair theorem. Such tests will be carried out by fitting EMRI models that include additional parameters that represent deviations from this assumption. We can thus assess the impact of calibration uncertainties on these tests of general relativity using the same Fisher matrix approach, applied to the extended parameter space that includes standard EMRI parameters, calibration uncertainty parameters and parameters characterising deviations from general relativity. We do this using the extended AK model described in~\citep{barackUsingLISAEMRI2007,babakScienceSpacebasedInterferometer2017}, which includes one additional parameter, $Q$, that represents an ``excess quadrupole'', i.e., a difference in the quadrupole moment of the spacetime relative to the value predicted from its mass and spin in the Kerr metric. In Figure~\ref{fig:EMRI_excess_quad} we show how the measurement precision of $Q$ varies as we increase the size of the calibration uncertainty. This was evaluated for a different EMRI system to the one considered above, with smaller mass for the inspiraling body, $\mu=10M_\odot$, smaller eccentricity ($e_{\rm pl} = 0.06$ at plunge) and a shorter observation time (1 year). The Fisher matrix was evaluated with the true excess quadrupole moment set to $Q=0$. The conclusions are not significantly affected by the specific choice of EMRI parameters. %\jon{Lorenzo please confirm} 
This quadrupole moment is again measured through its impact on the phase evolution of the binary and hence it is not surprising that the measurement precision behaves in a similar way to the other intrinsic parameters. We see a slight degradation in precision for large calibration uncertainties, but even with uncertainties of order unity, the measurement precision is within 5\% of its value in the absence of calibration error.

Based on this source, we conclude that EMRIs do not place a requirement on phase calibration uncertainty, but place a requirement on amplitude calibration uncertainty of $\sim 8\%$. Once again we expect this to be source specific, and this EMRI has signal-to-noise ratio of $30$, which is close to the threshold needed for EMRI detection. In the best case EMRIs could have signal-to-noise ratio of a few hundred, so we might expect to require amplitude calibration uncertainty of $\sim 1\%$ to ensure calibration has no impact for any individually resolved EMRI.
%Since intrinsic parameters determine the frequency evolution of the emitted gravitational wave signal, these are mostly affected by the phase calibration errors, as we can see from the left panel of Figure (\ref{fig:EMRI_calib}). In the range of the considered prior calibration uncertainty the impact of phase calibration errors is negligible since the precision of all the considered parameters does not degrade more than a factor of $2$. The most affected parameter is the initial eccentricity $e_0$, whose measurement precision degrades up to a factor of $\approx 1.25$. On the other hand, the amplitude calibration errors affect only the luminosity distance, whose measurement precision degrades up to a factor $2$ for $\sigma_C\approx 8\times 10^{-2}$.

%Furthermore, we study the impact of calibration errors on the ability of EMRIs to constrain the Kerrness of a black hole. We do so by considering the spacetime's mass quadrupole moment given by $Q$ as an additional parameter of the Analytical Kludge waveform model. The parameter $Q$ represents a phenomenological parametrization of hypothetical deviations from the general-relativistic quadrupole moment \citep{emri_Q_2007,babakScienceSpacebasedInterferometer2017} and therefore it can be used as an indication whether calibration errors can affect test of GR with EMRIs. As we can see from Figure~(\ref{fig:EMRI_excess_quad}), the parameter $Q$ is also unaffected by the presence of calibration errors.

%\lorenzo{add info on the excess quadrupole}

\subsection{Calibration requirements for inference on source populations}

All of the results described in this section have been based on measurements of the parameters of individual sources. LISA observations of many individual sources will also be combined to make statements about the properties of the underlying population and we also do not want population inference to be calibration limited. What happens when sources are combined will depend critically on whether the calibration errors for each source are independent of one another or not, which itself depends on how much the LISA calibration error might vary with time. With reference to Eq.~(\ref{eq:calmodSS}), it is important to know if $|1-C_i(f)| \ll |1-C(f)|$ for all $i$. If all sources are affected by the same calibration error then combining the sources can help to measure the calibration error and hence reduce uncertainties for individual events. This would rely on the fact that different sources have different frequency dependencies, and does not apply to an overall scaling, which would impact distance measurements and hence constraints on cosmological parameters. If individual sources have independent calibration errors then the impact on population parameters will at least partially average out as sources are combined and so the uncertainties in the population might well be comparable to those on the individual events. 

Without doing detailed simulations, we can place a conservative bound by considering a worst case scenario. In the absence of calibration errors we expect that the constraint on a population parameter improves like $1/\sqrt{N}$ as the number of observed sources, $N$, increases. In the worst case, the bias in the population parameter arising from calibration uncertainties will be common to all events and hence does not change as the number of events is increased. Therefore, we need the uncertainty arising from calibration error to be a factor of $1/\sqrt{N}$ smaller than the instrumental-noise induced uncertainty for a single event if we want to ensure that the calibration-error-induced uncertainty does not dominate after we combine $N$ events. Typical LISA source populations contain $O(100)$ events~\cite{eLISA:2013xep}. Thus, if the parameter uncertainties for individual events are increased by no more than 10\% when accounting for calibration uncertainties, we would expect population uncertainties to be no more than doubled. This corresponds to placing a threshold $\sigma_{\theta}/\sigma_{\theta0}<1.1$ on individual event parameter inference. Given the roughly linear growth of parameter uncertainties with the size of calibration uncertainties, we might expect to need a calibration requirement a factor of $10$ more stringent for population inference. Looking more carefully at the earlier results we find this constraint imposes amplitude calibration requirements of $10^{-2}$, $2 \times 10^{-2}$, $>1$ and $10^{-2}$ for population inference with galactic binaries, MBHBs, SBHBs and EMRIs respectively. The corresponding phase calibration requirements are $8\times10^{-3}$, $2\times10^{-3}$, $>1$ and $2\times10^{-2}$. Including an additional factor of a few to account for the fact that these sources may not be completely representative of the population, we conclude that if LISA amplitude calibration is better than a few$\times 0.1\%$ and phase calibration is better than $10^{-3}$ we would not expect to see any impact of calibration uncertainty on the LISA science output. 
%provided that the increase in parameter uncertainties from including calibration errors is no more than $\sim10\%$, we would not expect there to be any bias in population inference.
%\lorenzo{Does this mean that $\sigma_{\theta}/\sigma_{\theta0}<1.1$ ?}
%We saw for individual sources that calibration uncertainties begin to have an impact when they are of comparable magnitude to the precision with which the parameters can be measured in the absence of calibration uncertainty. We would therefore expect the calibration requirement for population inference to be no more than a factor of ten more stringent than the individual source requirement. Based on the earlier results, this requirement would be approximately $0.1\%$ in amplitude and a few $10^{-4}$ in phase, with the latter only being necessary for inference on massive black hole populations. \jon{Based on Fig 5. Need to discuss this further and have a clear message.}
%Therefore, if estimated LISA calibration uncertainties are smaller than $10\%$ of the acceptable uncertainties for individual source measurements, we would not anticipate there to be any bias in population inference. \jon{Add statement of what this means based on previous results??} 
If this more stringent requirement could not be met, then a more careful study would be required to understand if there would be any impact on population inference, which would have to account for possible time-variability of the LISA calibration error. This is beyond the scope of this paper.

\section{Using gravitational wave sources as calibrators}
\label{sec_GWcalibrators}
\subsection{Known galactic binaries as amplitude calibrators} \label{ref_GB_calibrator}
\begin{figure*}[h]
\includegraphics[width=\textwidth]{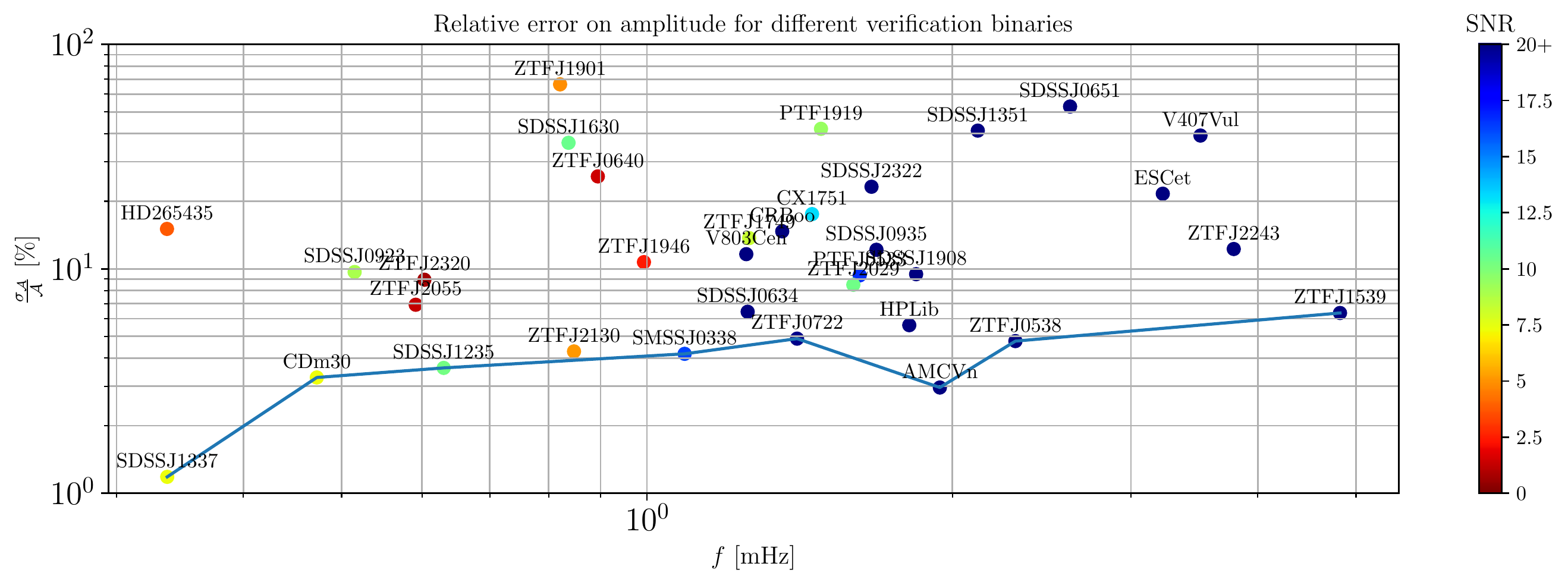}
\caption{Relative error on GW amplitude for different verification binaries as a function of the VGB central frequency. The SNR of each source is indicated by the color of the dot. The blue line shows the optimal choice of VGB to cover the largest frequency interval with the lowest relative error in amplitude and highest SNR.}%\stas{what is the orange line?}}
\label{fig:VGB_info}
\end{figure*}

For all source types we have seen that there is a degeneracy between measurement of distance and uncertainty in amplitude calibration, which ultimately limits our ability to measure distances. However, if distance is already known, the same effect can be used to directly measure the amplitude calibration error. Most of the sources LISA will observe will not have known distances, but the verification galactic binaries (VGBs) are an exception. These are galactic binaries which are already known from electromagnetic observations \cite{10.1093/mnras/sty1545}. There are a few tens of such systems for which the sky position is known near-perfectly, the period is known with a fractional uncertainty of a few $\times 10^{-6}$, the inclination is known to $\sim 10\%$ and the distance is known to between a percent and a few tens of percent. Combining these uncertainties, the GW amplitude in Eq.~(\ref{eq:GB_Amp}) is known to between $1\%$ and several tens of percent. Figure~\ref{fig:VGB_info} shows the GW amplitude uncertainties for all currently known VGBs. By comparing the observed amplitude of these sources to these expected amplitudes, we will be able to directly measure the amplitude calibration error as a function of frequency. The best VGBs for doing this (giving the tightest constraint over frequencies in the range $[0.3,5]$ mHz) are joined by a blue line in Figure~\ref{fig:VGB_info}. Unfortunately, the information on the initial phase of the VGBs is missing. Subsequently, we make the choice to consider the phase error as small in order to demonstrate the significant improvement of the calibration error parameters estimation when multiple VGBs are considered.

%Candidate verification binaries are a class of binary stars with ultrashort orbital periods, consisting of a white dwarf or neutron star primary and a compact Helium-star/white dwarf/neutron star secondary.
%These observations provide a lot of information that will help to refine the parameter estimation during the detection of the gravitational wave. So far, the a-priori available on many sources cover the position in the sky (with a near perfect precision on the right ascension and declination, and a precision of the order of 10\% on the inclination), the distance (with a relative error of the order of percent), the period (with a relative error of the order of a few parts per million). The amplitude of the gravitational wave, defined in equation (\ref{eq:GB_Amp}), is known to within 1\% for the best sources. No information is available on the initial phase. The signal-to-noise ratio (SNR) of some of these sources can reach a value of 200 after 4 years, making them perfect calibrators that will allow to extract part of the coefficients of the calibration error. A well chosen set of sources (shown with the blue line in figure \ref{fig:VGB_info}) allows to cover the frequency interval.

To illustrate this procedure, we simulate the observation of a set of 8 VGBs and then recover posterior distributions on the source amplitude and the amplitude calibration spline parameters, keeping the other sources parameters fixed to their true values. We impose Gaussian priors on the weights at each knot of the calibration spline, with width of $0.1$ in strain, and we consider the case either that there is no prior electromagnetic information on the source distance, or that we use EM prior information from the best 8 VGBs identified in Figure~\ref{fig:VGB_info}. The waveforms were generated for a 4 year LISA observation, using a LISA response based on analytical orbits and computed using second order time delay interferometry (TDI2.0), and with the instrumental noise generated and modelled according to the first version of the Scientific Requirement Document (SciRDv1). We include a calibration error in the simulated data, drawn randomly from the assumed prior distribution.

\begin{figure*}
\centering
\begin{subfigure}{.5\textwidth}
  \centering
  \includegraphics[width=\textwidth]{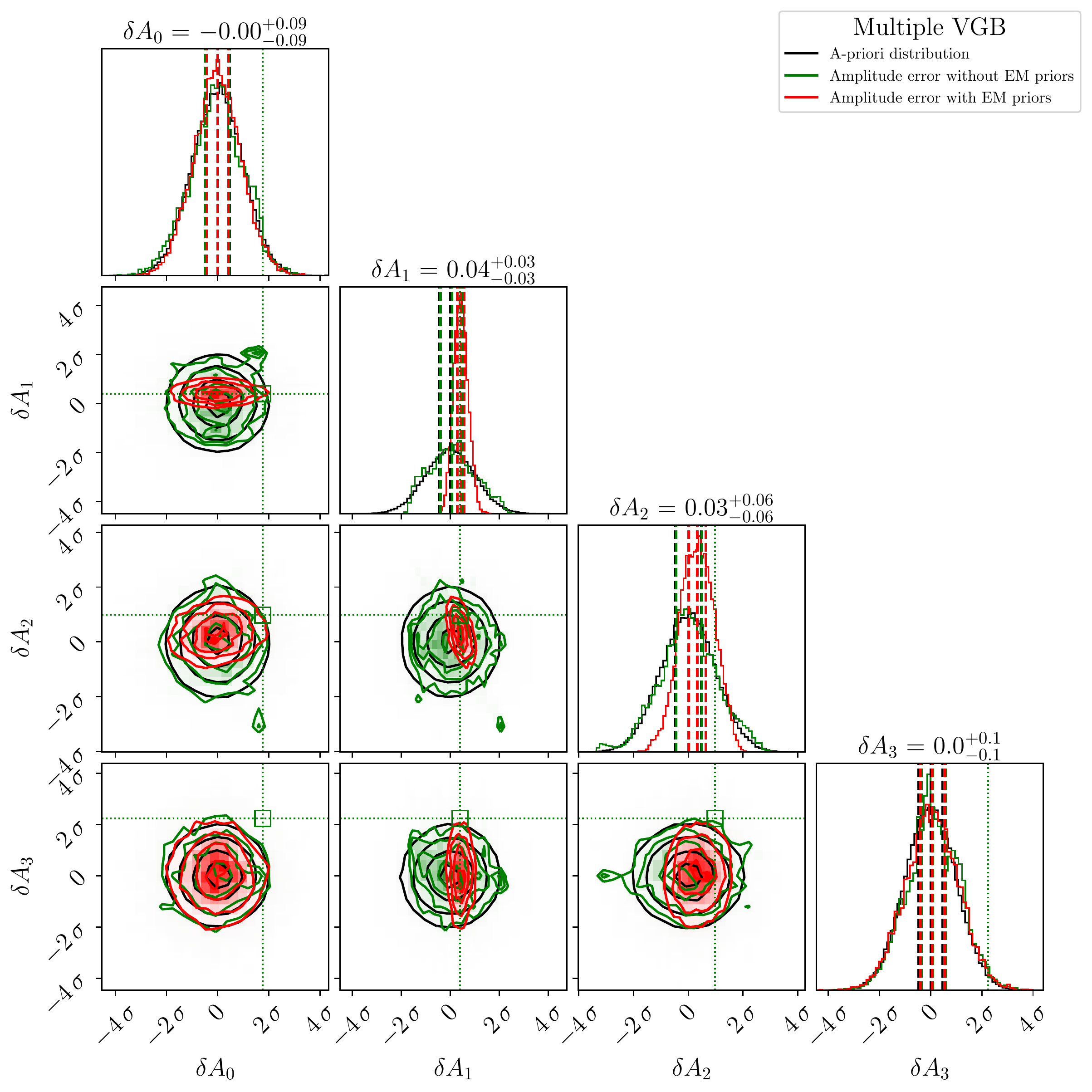}
  \caption{Amplitude}
  \label{fig:VGB_Amp}
\end{subfigure}%
\begin{subfigure}{.5\textwidth}
  \centering
  \includegraphics[width=\textwidth]{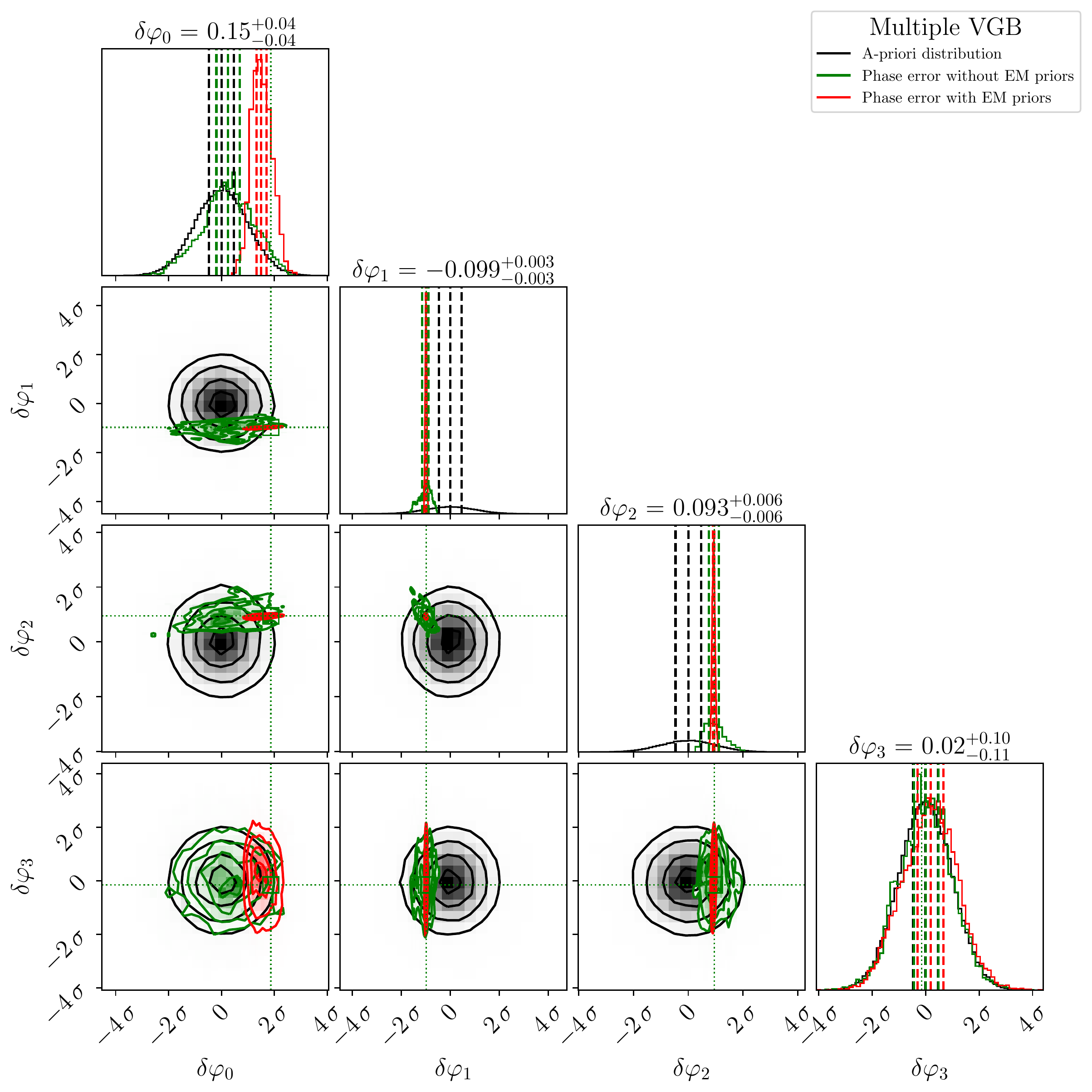}
  \caption{Initial phase}
  \label{fig:VGB_Pha}
  %\etienne{Change the calibration error parameters notation}
\end{subfigure}
\caption{Posterior distribution for the calibration error parameters in three scenarios : without any calibration error (black), with the calibration error without (green) and with (red) the electromagnetic counterparts a-priori. Left corresponds to the amplitude error parameters ($\delta A_0,\delta A_1,\delta A_2,\delta A_3$) while right corresponds to the phase error parameters ($\delta \varphi_0,\delta \varphi_1,\delta \varphi_2,\delta \varphi_3$). The dashed green lines indicate the injected value of the calibration error, which was draw randomly from the assumed prior.}
\label{fig:test}
\end{figure*}

The results are shown in the left-hand panel of Figure~\ref{fig:VGB_Amp}, along with the prior distribution on the calibration parameters for comparison. We see that in the absence of a prior from  the EM data, the posterior recovered for the amplitude calibration parameters coincides with the prior, as we might expect. We cannot decouple the unknown distance from the unknown amplitude calibration without additional information. However, once the EM prior is added, the posterior on $\delta A_1$ and $\delta A_2$ are narrower than the prior distribution, indicating that we are successfully measuring these quantities.  These are the weights of the amplitude calibration spline at $f=1$mHz and $f=10$mHz.  All the VGBs lie at frequencies between these two knots, so it makes sense that this is where the amplitude calibration uncertainty can be measured. We see that VGBs could provide amplitude calibration at the level of a few percent.

GBs also show a similar degeneracy between the phase offset of the GW signal and the phase calibration uncertainty, and so we might hope to use EM measurements of the former to determine the latter. Unfortunately, the VGB phase offset is not normally well constrained by EM observations. However, to illustrate the possibility, we assume that we do have an EM prior on $\phi_0$ and repeat the above procedure, now fitting for the system phase and the four phase calibration parameters, and imposing a Gaussian prior on the latter with width of $0.1$ radians. These results are shown in the right-hand panel of Figure~\ref{fig:VGB_Amp}. In this case, the degeneracy between $\varphi_0$ and the calibration parameters is not complete and so even in the absence of EM information the posterior on the phase calibration parameters, in the region where the signals lie, is different to the prior. This will be discussed further for EMRIs in the next subsection. When the EM prior is added the constraints on all the phase calibration parameters becomes tighter, with measurement precisions less than one hundredth of a radian.

We include that EM observations of VGBs can be used to provide information on calibration of the LISA instrument. This calibration will be limited to the frequency range where the VGBs lie, which is $0.1$ to $5$ mHz, and amplitude calibration will be limited by how well distances to the VGBs are known, roughly $1\%$ at present. The publication of the third edition of the GAIA catalog \url{https://www.cosmos.esa.int/web/gaia/earlydr3}, expected soon, should significantly improve distance measurements of several VGBs and provide new VGB candidates. By the time LISA flies we might therefore be in a position to do somewhat better than this, although it is unlikely we will meet the necessary amplitude calibration precision of $\sim 0.1\%$ identified above. Nonetheless, it is reassuring to know that even if the calibration of LISA was completely unknown we could use this procedure to minimize the impact on LISA science objectives.

\subsection{Inspiraling binaries as phase calibrators}
%--------------------------------------------------------------------
The previous analysis relied on using additional EM information to calibrate the LISA data stream. However, we have also seen that for many of the source types, parameter uncertainties do not grow without bound as the phase calibration uncertainties are made arbitrarily large. This is because the way in which (the assumed model of) phase calibration errors affects the waveform phase is distinct and distinguishable from the phase evolution of the signal. This means that it should be possible to measure both simultaneously from the data, ``piggy-backing'' the measurement of the calibration uncertainty onto the measurement of the astrophysical signal. We explore this here, using EMRI systems, which show the most limited degradation in parameter measurements of all the sources considered here. EMRIs are expected to be on eccentric and inclined orbits, so their GW signals contain a large number of significant harmonic modes. These evolve slightly differently to one another, but are all affected by the calibration error in the same way, which is likely what facilitates separation of the two effects.

\begin{figure*}
    \centering
    \includegraphics[width=\textwidth]{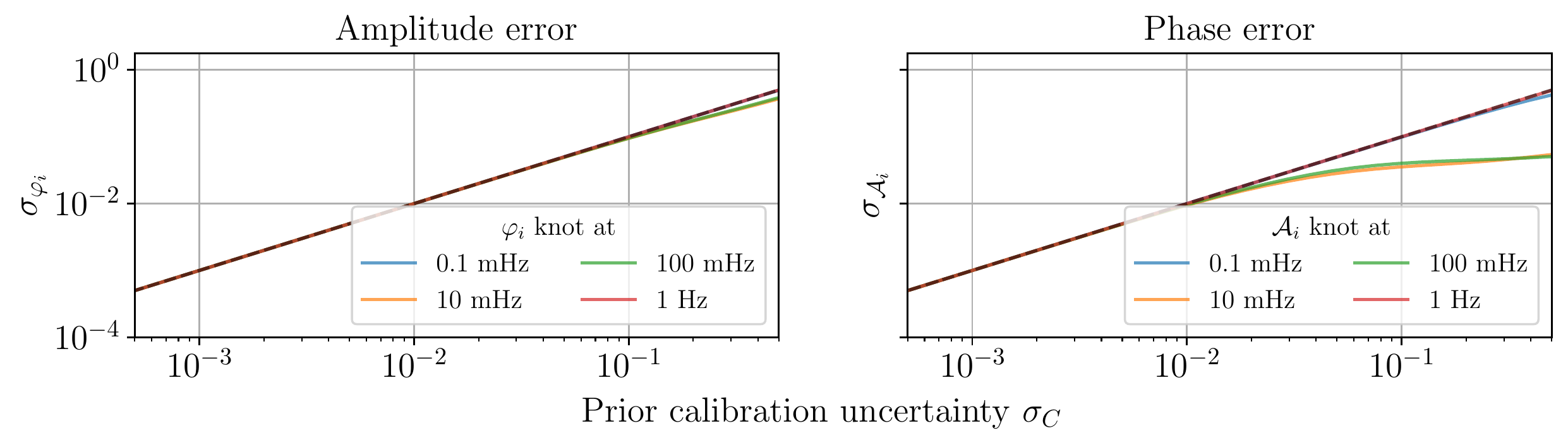}
    \caption{
        Estimated uncertainty $\sigma_\theta$ of the calibration parameters $\delta \varphi_i, \delta A_i$ as a function of the prior calibration uncertainty $\sigma_C$, when fitting these parameters simultaneously with those of the EMRI signal represented in Figure~\ref{fig:EMRI_calib}. The black dashed diagonal line shows the case where the estimated uncertainty $\sigma_\theta$ equals the prior calibration uncertainty $\sigma_C$. Deviations from this line indicate that we are able to constrain the calibration parameters $\delta \varphi_i, \delta A_i$ using the observation.}
%        \lorenzo{make plots with same axis and include information on the system and change label of the calibration parameters} \jon{All other plots have amplitude left and phase right. Swap round?}\stas{shouldn't we use $\mu$ instead of $\theta$ here?}
    \label{fig:EMRI_as_calibrators}
\end{figure*}

To investigate this, we repeat the same Fisher matrix analysis as before, but now look at the other diagonal elements of the inverse Fisher matrix, which correspond to the calibration error model parameters, $\{\delta \phi_i, \delta A_i\}$. These elements of the inverse Fisher matrix characterise how well we can measure the calibration parameters. If they are comparable to the assumed prior width, $\sigma_C$, then we have not gained any information from fitting these parameters at the same time as the source. If the posterior is narrower than $\sigma_C$ then it shows we have measured the calibration uncertainty. We show these results in Figure~\ref{fig:EMRI_as_calibrators}, as a function of the assumed calibration uncertainty. We see that EMRIs cannot be used to measure the amplitude calibration uncertainty. This is to be expected as that degeneracy is perfect. Only by adding external information, as described in the previous subsection, can we hope to measure the amplitude calibration. However, we see that we can use EMRI observations to measure the phase calibration. Once the phase calibration uncertainty is greater than $0.01$ radians the measured uncertainty starts to flatten off, showing that it is being determined from the data. It only flattens off for $\delta\phi_1$ and $\delta\phi_2$, but that is because these are the parameters that affect the calibration in the frequency range where the EMRI is generating GWs. We conclude that we will be able to measure the phase calibration uncertainty to better than a few times $10^{-2}$ radians, by fitting it simultaneously with the parameters of our sources. Once again, this is not quite at the level we argued was required to ensure no impact on science, but it is not significantly worse. So even under the most pessimistic assumptions about calibration uncertainties, LISA should still be able to produce high quality scientific results.

\section{Summary}
\label{sec_summary}
We have studied the impact of calibration uncertainties, i.e., differences between the true signal component of a data set and the model used to describe that signal component in data analysis, on LISA science output. We have focused on the impact of calibration uncertainties on parameter estimation for individual sources, and have considered all the major source types that LISA is expected to observe. These results show that amplitude calibration uncertainties directly affect our ability to measure the distances to all types of gravitational wave source. Amplitude calibration uncertainty becomes limiting when it becomes comparable to the typical precision with which amplitudes can be measured, which is typically $1\%$.  %but can be as small as $0.1\%$ for the loudest massive black hole binary sources. 
In a similar way, phase calibration uncertainties directly limit our ability to measure the absolute phase of binaries, but the initial phase is typically not a parameter that conveys a significant amount of astrophysical information. Phase calibration uncertainties also affect our ability to measure other physical parameters of the system, such as masses and spins, that determine the phase evolution of the gravitational waveform. However, in that case the impact of calibration uncertainty is limited and even large phase calibration uncertainties at worst degrade the determination of the system parameters by a factor of $2$.

The impact of phase calibration uncertainties is limited because the slowly-varying (as a function of frequency) calibration error that we assume here is not degenerate with the more rapidly varying phase of a typical gravitational wave source. This effect allows us to simultaneously fit for the gravitational wave signal and the calibration error, with the gravitational wave acting as a carrier beat note for measuring the calibration uncertainty. Using extreme-mass-ratio inspirals we find that phase calibration errors of the level of a few percent could be measured. Amplitude calibration uncertainty is degenerate with the unknown distance to the source and so cannot be measured in the same way. However, some galactic binary sources have distances measured as a percent level through electromagnetic observations. These verification binaries could be used to measure amplitude calibration uncertainties with a precision of a few percent.

We have focused on the impact of calibration uncertainties on parameter estimation in this paper, because it is our ability to measure the parameters of sources with high precision that drives most of the scientific objectives of the LISA mission~\cite{eLISA:2013xep}. This is not the only thing that calibration uncertainties will impact, however. Calibration errors will also impact our ability to detect sources in the data, because mismatches between the detection templates and the signal component of the data will lead to losses in signal-to-noise ratio. For the size of calibration uncertainties that are limiting for parameter estimation, this effect will be at the level of tens of percent of the total number of events. This is much smaller than the intrinsic uncertainties in the astrophysical rates of LISA events, and uncertainties arising from weak lensing of sources at cosmological distances. Additionally, the observed signal-to-noise ratio will fluctuate due to uncertainties in the noise spectral density of the detector, and these uncertainties are expected to be much larger than those in the calibration model. Calibration uncertainties will also affect our ability to infer the parameters that characterise the astrophysical population of sources, obtained by combining observations of multiple sources hierarchically. To fully understand the impact of calibration uncertainties on population inference, we would need to model the time dependence of the calibration error, making assumptions about the degree of correlation between the calibration uncertainty in different events. For the moment, we can crudely estimate that calibration uncertainties would have to be about one order of magnitude smaller to be sure they would not limit typical LISA population inference studies. Finally, we have not considered the impact of calibration uncertainties on the characterisation of stochastic gravitational wave backgrounds, as these require a different analysis approach to the one described here. However, phase calibration uncertainties should not affect our ability to characterise stochastic backgrounds, as the expected phase is random. Amplitude calibration uncertainties would be degenerate with stochastic background spectra that had a similar form, but the spline model assumed here does not resemble the power-law form of a typical stochastic or cosmological background, and the calibration uncertainty would be significantly sub-dominant to uncertainties arising from lack of knowledge of the spectral density of the instrumental noise.

The results in this paper give a baseline target for the calibration of the LISA instrument, but the analysis could be extended in a number of ways if necessary. Firstly, we have used a very slowly varying model for the calibration uncertainty. This was well motivated from instrumental considerations, but it does mean that the calibration uncertainty is not very degenerate with the gravitational waveform, and this is one of the reasons that phase uncertainties are not limiting. Increasing the number of spline points, or adopting an alternative model such as representing the uncertainty as a realisation of a Gaussian process, would provide the flexibility to model more rapidly varying uncertainties. Secondly, we have taken a single calibration spline to represent the whole data set, while in reality the calibration might be expected to vary over time. This could have an impact on the determination of sky position, which relies on the variation in amplitude over the course of a year. We have also ignored realistic features of the instrumental data, such as data gaps. The assumed model of the time-dependence of the calibration error will influence conclusions about population inference in particular. Finally, in this work we have only placed limits on the calibration uncertainty, but LISA science will also be limited by our degree of knowledge of the noise model. Understanding the impact of and placing requirements on our noise-model uncertainty is also important to ensure the mission delivers on its high scientific promise. %\stas{Should we mention that the systematic error in GW model could be added in somewhat similar way?} \jon{Not sure we need to, unless we wanted to have a generic statement about the fact that this methodology is useful to other scientific questions.}

%other impacts of calibration
%outstanding issues - claibration model assumptions, time dependence etc.

\begin{appendix}
\section{Setting calibration requirements based on induced parameter errors}
\label{sec_bias}
%\responsible{Jon}
The Fisher matrix-based approach that we use in this paper implicitly assumes that the calibration model is being fitted for at the same time as the individual sources when the data is analysed. Another approach to setting calibration requirements is to ask how large the uncorrected calibration error can be before it starts to lead to biases in the recovered parameters.

The bias can be estimated in the linear signal approximation using the formula
\begin{align}
%\Delta \theta^i &= ((\Gamma^{\theta\theta})^{-1})_{ij} (\partial_j h | \sum_k ((\mathbf{C}(f) - 1) \mathbf{h} (\vec\theta_k)).
\Delta \theta_{\rm b}^i = ((\Gamma^{\theta\theta})^{-1})_{ij} (\partial_j h | ((\mathbf{C}(f) - 1) \mathbf{h} (\vec\theta)).
\label{eq:CVBias}
\end{align}
This expression can be used to assess whether a particular calibration error, described by $\mathbf{C}(f)$, would lead to significant biases in parameter estimation, by comparing the calibration error-induced parameter uncertainty, $\Delta \theta_{\rm b}^i$,  to the corresponding statistical uncertainty arising from instrumental noise, which is given by the square root of the corresponding diagonal element of the inverse Fisher Matrix, $\sqrt{(\Gamma^{\theta\theta})^{-1}_{ii}}$.

This approach can be related to the joint Fisher Matrix approach adopted here by working in the linear signal approximation. We assume that the calibration model parameters are close to the reference value of zero and write
$$
\mathbf{C}(f) - 1 \approx \frac{\partial \mathbf{C}}{\partial \mu_k} \Delta \mu_k.
$$
The prior that the calibration errors are centred around zero with uncertainties characterised by $\Sigma^{\mu \mu}$ is equivalent to the assumption that $\Delta \mu_k \sim N(0, (\Sigma^{\mu\mu})^{-1})$. As the precise calibration error is unknown, the parameter uncertainty induced by the calibration error
$$
\Delta \theta_{\rm b}^i = ((\Gamma^{\theta\theta})^{-1})_{ij} (\partial_j h | \partial_k \mathbf{C} \; \mathbf{h} (\vec\theta)) \Delta \mu_k
$$
is a random variable. Here we are using the notation $\partial_k \mathbf{C} = \partial \mathbf{C}/\partial \mu_k$. This random variable has zero mean and variance
\begin{widetext}
\begin{align}
    \langle \Delta \theta_{\rm b}^i \; \Delta \theta_{\rm b}^j\rangle &=  ((\Gamma^{\theta\theta})^{-1})_{il} (\partial_l h | \partial_k \mathbf{C} \; \mathbf{h} (\vec\theta)) \langle \Delta \mu_k \Delta \mu_m \rangle ((\Gamma^{\theta\theta})^{-1})_{jn} (\partial_n h | \partial_m \mathbf{C} \; \mathbf{h} (\vec\theta)) \nonumber \\
    &=((\Gamma^{\theta\theta})^{-1})_{il} 
    (\Gamma^{\theta\mu})_{lk} (\Sigma^{\mu\mu})^{-1}_{km} ((\Gamma^{\theta\theta})^{-1})_{jn} (\Gamma^{\theta\mu})_{nm}.
\end{align}
\end{widetext}
As this bias is independent of the noise-induced statistical uncertainty, the total variance in the parameter uncertainties is the sum of the statistical and systematic variances
\begin{align}
 \langle \Delta \theta^i \; \Delta \theta^j\rangle = (\Gamma^{\theta\theta})^{-1}_{ij} + \nonumber \\
  ((\Gamma^{\theta\theta})^{-1}
    (\Gamma^{\theta\mu}) (\Sigma^{\mu\mu})^{-1} (\Gamma^{\theta\mu})^{T} ((\Gamma^{\theta\theta})^{-1})^T)_{ij}.
    \label{eq:BiasVarTot}
\end{align}
We can compare this to the expression used here, based on the inverse of the joint Fisher Matrix, Eq.~(\ref{eq:FMinvtheta}), which can also be expressed in the alternative form
\begin{eqnarray}
(\Gamma^{\theta\theta})^{-1} +
(\Gamma^{\theta\theta})^{-1} \Gamma^{\theta\mu} \left(\Gamma^{\mu\mu} + \Sigma^{\mu\mu} -
\right. \nonumber \\
\left. (\Gamma^{\theta\mu})^T (\Gamma^{\theta\theta})^{-1} \Gamma^{\theta\mu}\right)^{-1} (\Gamma^{\theta\mu})^T (\Gamma^{\theta\theta})^{-1}.
\end{eqnarray}
We see that this has a similar form to expression~(\ref{eq:BiasVarTot}), but we have replaced $\Sigma^{\mu\mu}$ by $ \Sigma^{\mu\mu} + \Gamma^{\mu\mu} - (\Gamma^{\theta\mu})^T (\Gamma^{\theta\theta})^{-1} \Gamma^{\theta\mu}$. This difference reflects the impact of fitting for the calibration model. The inverse of this term can be interpreted as the residual variance (or uncertainty) in the calibration model, which is reduced when it is fitted out of the data.

%Conclusions for the biases on source parameters arising from miscalibration of the source in question are likely to be comparable to those estimated from the Fisher matrix. However, it will be of interests and importance to understand the effect of miscalibration of loud sources, or a population of very many quieter sources, on PE for other signals.

As a final remark, we note that Eq.~(\ref{eq:CVBias}) can also be used to assess our tolerance to unmodelled calibration effects. In practice, we will fit for some kind of calibration uncertainty model at the same time as fitting for the parameters of the sources present in the data. Any component of the true calibration uncertainty that can not be represented by that model will remain in the data and lead to biases. These biases can be estimated using the above formula, but with $(\mathbf{C}(f) - 1)$ replaced by the residual calibration error $\mathbf{C}_{\rm true}(f; \mu) - \mathbf{C}_{\rm mod}(f; \mu)$ and the set of parameters and Fisher Matrix expanded to include the calibration model parameters. We will not consider this further here, but such biases could be assessed once numerical calibration uncertainty simulations from instrumental modelling are available.
%TO HERE

\section{Cross-checks of the Fisher matrix results} \label{sec_analyticalfisher}
The Fisher matrix results presented in this paper were computed numerically. I this appendix we report a number of cross-checks that were used to verify the validity of these results. We derive the Fisher matrix for Galactic binaries analytically ad that for SBHBs semi-analytically. We also verify the GB results by directly evaluating a posterior on the calibration parameters using MCMC methods. To facilitate the analytic calculations, we first introduce an alternative model for the calibration error.

\subsection{Calibration error knots formulae} \label{sec_cubicparameters}
In the main part of this paper, the calibration error is described using two cubic splines. In order to verify the validity and stability of the Fisher matrix, we will use in this appendix a calibration error that is just a simple cubic to simplify the analytical derivation. While the cubic spline model has considerably more flexibility in general, when using only four spline knots it is not so different to a cubic, and using the latter as an alternative model allows us to derive analytically (respectively semi-analytically) the Fisher matrix for a GB (resp. SBHB). %\jon{Between two knots a cubic spline is a cubic. So you don't need to use a different model to derive these results, particularly for GBs for which the entire signal will typically be between two knots.} \etienne{For the case of SBHBs the derivation is necessary (or it at least simplifies the calculation compared to the interpolated cubic) but you are totally right in the case of GBs where the two definitions are equivalent.}

The amplitude (respectively phase) error function is a cubic that is defined with 4 parameters $\{\delta A_i\}$ (resp. $\{\delta \varphi_i\}$) defined at specific knots $\{ 0.1, 1, 10, 1000\}$. The total calibration error can be written:
\begin{eqnarray}
C(f,\{\delta A_i,\delta \phi_i\}) &=& \left(1+\sum_{n=0}^{3} a_n(\{\delta A_i\}) \left(\log_{10} f\right)^n \right) \times \nonumber \\ 
& &
e^{2\pi i \sum_{n=0}^{3} p_n(\{\delta \varphi_i\}) \left(\log_{10} f\right)^n}\, ,
\label{eq_Cf_cubic}
\end{eqnarray}
where $\{a_i\}$ and $\{p_i\}$ corresponds to the cubic function parameters and are linear combination of $\{\delta A_i\}$ and $\{\delta \varphi_i\}$ 

Using $\delta X(f)$ to denote either the amplitude or the phase error function we have:
\begin{equation}
\tilde{\delta X}(f) = x_0 + x_1 \left(\log_{10} f\right) + x_2 \left(\log_{10} f\right)^2 + x_3 \left(\log_{10} f\right)^3
\end{equation}
where $\{x_i\}$ are the cubic spline parameters, which relate to the weights at the knots, $\{\delta X_i\}$, via 
\begin{widetext}
\begin{equation}
\left\{
\begin{array}{c}
\delta X_0 = \delta X(f_0) = x_0 + x_1 \left(\log_{10} f_0\right) + x_2 \left(\log_{10} f_0\right)^2 + x_3 \left(\log_{10} f_0\right)^3\\
\delta X_1 = \delta X(f_1) = x_0 + x_1 \left(\log_{10} f_1\right) + x_2 \left(\log_{10} f_1\right)^2 + x_3 \left(\log_{10} f_1\right)^3\\
\delta X_2 = \delta X(f_2) = x_0 + x_1 \left(\log_{10} f_2\right) + x_2 \left(\log_{10} f_2\right)^2 + x_3 \left(\log_{10} f_2\right)^3\\
\delta X_3 = \delta X(f_3) = x_0 + x_1 \left(\log_{10} f_3\right) + x_2 \left(\log_{10} f_3\right)^2 + x_3 \left(\log_{10} f_3\right)^3\\
\end{array}
\right. .
\end{equation}
\end{widetext}
In our case, $f_0 = 10^{-4}$, $f_1 = 10^{-3}$, $f_2 = 10^{-2}$ and $f_3 = 1$ Hz. Writing these equations in matrix form, we can readily invert them to obtain the relationship between the cubic spline parameters and the error parameters :
\begin{widetext}
\begin{equation}
\left(
\begin{array}{c}
\delta X_0 \\
\delta X_1 \\
\delta X_2 \\
\delta X_3
\end{array}
\right)
=
\left(
\begin{array}{c c c c}
1 & -4 & 16 & -64 \\
1 & -3 & 9 & -27 \\
1 & -2 & 4 & -8 \\
1 & 0 & 0 & 0 
\end{array}
\right)
\left(
\begin{array}{c}
 x_0 \\
 x_1 \\
 x_2 \\
 x_3 
\end{array}
\right)
\Rightarrow
\left(
\begin{array}{c}
 x_0 \\
 x_1 \\
 x_2 \\
 x_3 
\end{array}
\right) =
\left(
\begin{array}{cccc}
 0 & 0 & 0 & 1 \\
 -\frac{3}{4} & \frac{8}{3} & -3 & \frac{13}{12} \\
 -\frac{5}{8} & 2 & -\frac{7}{4} & \frac{3}{8} \\
 -\frac{1}{8} & \frac{1}{3} & -\frac{1}{4} & \frac{1}{24}
\end{array}
\right)
\left(
\begin{array}{c}
\delta X_0 \\
\delta X_1 \\
\delta X_2 \\
\delta X_3
\end{array}
\right) .
\end{equation}
\end{widetext}
\subsection{Galactic binary analytical confirmation} \label{subsec:GB_ana}
Here we recompute the Fisher matrix for GBs analytically. The noiseless TDI combinations (here $X$) used to simulate the gravitational wave can be represented with the following formula :
\begin{equation}
\tilde{d}(f) \equiv X(f,\theta,\mu) = C(f,\mu) R_{X}(f,\theta) \mathcal{A} e^{i \Phi(f,\theta)}
\end{equation}
where $\theta$ are the waveform parameters, $\mu$ the calibration error parameters, $C$ is the calibration error function, $R_X$ is the LISA transfer function for the TDI $X$ combination, $\mathcal{A}$ is the waveform amplitude and $\Phi(t)$ is the waveform phase defined in equation (\ref{eq:GB_WF}). For GBs the LISA transfer function $R_{X}$ only depends on the sky location and is independent from the intrinsic waveform parameters.

In the main part of the paper, we numerically simulated the 3 TDI combinations and then calculated the Fisher matrix via equation (\ref{eq:fisher_def}). To do this, it is necessary to calculate the derivative of each TDI combination with respect to the different parameters (waveform + calibration). Since the calibration error only affects the amplitude and initial phase of the gravitational wave (as seen in Figure \ref{fig:GB_FIM}), we will concentrate on these waveform parameters only :
\begin{equation}
\begin{aligned}
\frac{ \mathrm{d} X(f,\theta,\mu)}{\mathrm{d} \mathcal{A}} =& \frac{1}{\mathcal{A}} X(f,\theta,\mu) \\
\frac{ \mathrm{d} X(f,\theta,\mu)}{\mathrm{d} \mathcal{\phi_0}} =& - i X(f,\theta,\mu) \\
\frac{ \mathrm{d} X(f,\theta,\mu)}{\mathrm{d} a_n} =& \left(\log_{10}(f)\right)^n X(f,\theta,\mu) \\
\frac{ \mathrm{d} X(f,\theta,\mu)}{\mathrm{d} p_n} =& 2 \pi i \left(\log_{10}(f)\right)^n X(f,\theta,\mu) \\
\end{aligned}
\label{eq:GB_derivative}
\end{equation}
where we used the calibration error function defined in equation (\ref{eq_Cf_cubic}) and we evaluate the Fisher matrix for zero calibration error, to ensure no bias, as discussed in appendix~\ref{sec_bias}. %\jon{Changed Eq. reference to point to cubic model since I think this was what was used here, but please cofirm.}

As the source is a galactic binary, we can assume the waveform frequency is equal to the central frequency of the GB ($f \to f_0$). Using this transformation, the definition of the inner product integral is simplified : 
\begin{equation}
%(\mathbf{a}|\mathbf{b}) 
(a|b) \equiv 4 \mbox{Re} \left[\frac{\tilde{a}^*(f_0) \tilde{b}(f_0)}{S_n(f_0)}\right] \, \Delta f
\end{equation}
where $\Delta f$ is the Fourier transform resolution of the $\tilde{a}$ and $\tilde{b}$ quantities.

The non zero Fisher matrix elements for the waveform amplitude, waveform initial phase and calibration error parameters can thus be analytically calculated :
%\begin{equation}
%\tiny
%\Gamma = \text{SNR}^2 \left(
%\begin{array}{cccccccccc}
%\frac{1}{\mathcal{A}^2} & 0 & \frac{1}{\mathcal{A}} & \frac{\log_{10}(f_0)}{\mathcal{A}} & \frac{\left(\log_{10}(f_0)\right)^2}{\mathcal{A}} & \frac{\left(\log_{10}(f_0)\right)^3}{\mathcal{A}} & 0 & 0 & 0 & 0 \\
%& -1 & 0 & 0 & 0 & 0 & 2\pi & - 2 \pi \log_{10}(f_0) & 2\pi \left(\log_{10}(f_0)\right)^2 & - 2 \pi \left(\log_{10}(f_0)\right)^3 \\
%& & 1 & \log_{10}(f_0) & \left(\log_{10}(f_0)\right)^2 & \left(\log_{10}(f_0)\right)^3 & 0 & 0 & 0 & 0 \\
%& & & \left(\log_{10}(f_0)\right)^2 & \left(\log_{10}(f_0)\right)^3 & \left(\log_{10}(f_0)\right)^4 & 0 & 0 & 0 & 0 \\
%& & &  & \left(\log_{10}(f_0)\right)^4 & \left(\log_{10}(f_0)\right)^5 & 0 & 0 & 0 & 0 \\
%& & &  &  & \left(\log_{10}(f_0)\right)^6 & 0 & 0 & 0 & 0 \\
% & & & & & & 4 \pi^2 & 4 \pi^2\log_{10}(f_0) & 4 \pi^2\left(\log_{10}(f_0)\right)^2 & 4 \pi^2\left(\log_{10}(f_0)\right)^3\\
% & & & & & & & 4 \pi^2\left(\log_{10}(f_0)\right)^2 & 4 \pi^2\left(\log_{10}(f_0)\right)^3 & 4 \pi^2\left(\log_{10}(f_0)\right)^4\\
% & & & & & & &  & 4 \pi^2\left(\log_{10}(f_0)\right)^4 & 4 \pi^2\left(\log_{10}(f_0)\right)^5\\
% & & & & & & &  &  & 4 \pi^2\left(\log_{10}(f_0)\right)^6\\
%\end{array}
%\right)
%\end{equation}
\begin{equation}
\begin{aligned}
\Gamma^{\mathcal{A},\mathcal{A}} &= \left(\frac{\rho}{\mathcal{A}}\right)^2, \quad \Gamma^{\phi,a_n} = \rho^2 \frac{\log_{10}(f)^{n}}{\mathcal{A}},  \quad \Gamma^{\phi,\phi} = -\rho^2, \\ 
\Gamma^{\phi,p_n} &= 2 \pi \rho^2 \left(-\log_{10}(f)\right)^{n}, 
\Gamma^{a_n,a_m} = \rho^2 \left(\log_{10}(f)\right)^{n+m},\\
\Gamma^{p_n,p_m} &= 4 \pi^2 \rho^2 \left(\log_{10}(f)\right)^{n+m}
\end{aligned}
\end{equation}
where we have defined the signal-to-noise ratio via:
\begin{equation}
\rho = 4 X(f_0)^2/S_n(f_0) \Delta f\,.
\end{equation}

\begin{figure*}
    \centering
    \includegraphics[width=\textwidth]{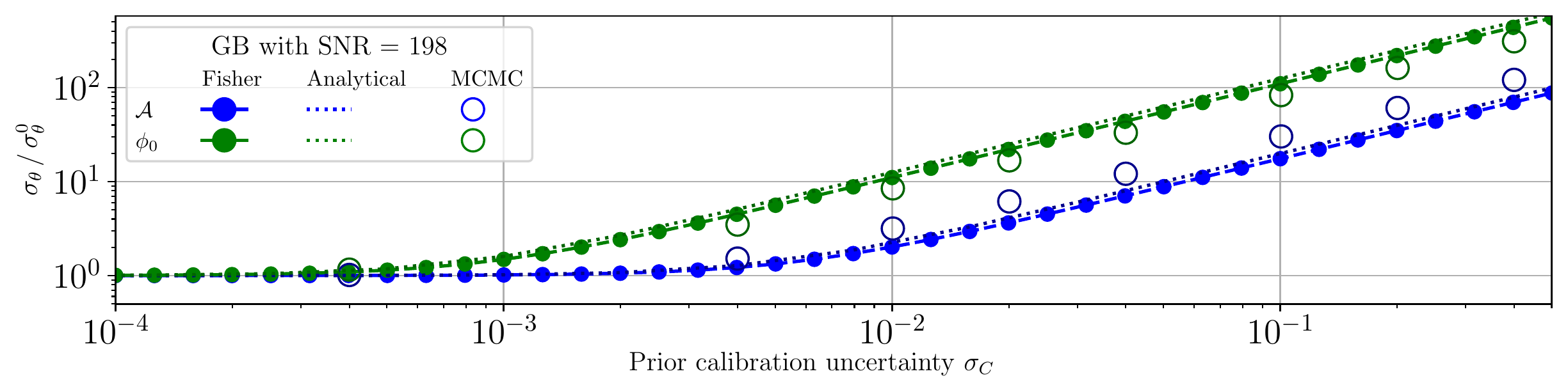}
    \caption{Ratio between the uncertainty of a waveform parameter with ($\sigma_{\theta}$) and without (${\sigma_{\theta}}_0$) calibration error uncertainty as a function of the calibration error uncertainty $\sigma_C$. The dotted lines represent the numerical  Fisher matrix results, which are in full agreement with the analytical predictions shown with solid lines.} %\etienne{I need to add the results from the MCMC}
    \label{fig:GB_ana}
\end{figure*}

From equation (\ref{eq:FMinvtheta}), we can extract the ``calibrated'' waveform amplitude and initial phase uncertainties (index $_c$) compared to the nominal uncertainties in the absence of calibration uncertainty (index $_0$) :
\begin{equation}
\begin{aligned}
\sigma_{\mathcal{A}_c}^2 &= \sigma_{\mathcal{A}_0}^{2} + \mathcal{A}_0^2
\sum_{n} \left(\log_{10}(f_0) \right)^{2n} \sigma_{a_n}^2 \\
\sigma_{\phi_c}^2 &= \sigma_{\phi_0}^{2} + 4 \pi^2 \sum_{n} \left(\log_{10}(f_0) \right)^{2n} \sigma_{p_n}^2
\end{aligned}
\end{equation}
where $\sigma_{a_n}$ and $\sigma_{p_n}$ are the calibration error spline-function parameters defined in equation (\ref{eq_Cf_cubic}). %\jon{Ditto} 
These analytical predictions are shown along with the numerical results in Figure~\ref{fig:GB_ana} and we see they are in full agreement.

\subsection{Galactic binary MCMC confirmation} \label{subsec:GB_MCMC}
The Fisher matrix is an approximation to the shape of the posterior, valid in the limit of high signal-to-noise ratio. To verify that it is valid for the calculations being performed here, we will now cross-check against results based on the numerical evaluation of the full Bayesian posterior distribution computed via Markov Chain Monte Carlo (MCMC) sampling.  %\jon{Is it correct you used MCMC for this, as well as monte carloing over realisations of the calibration error and noise?} \etienne{Correct. I could have used a fully MCMC sampling instead of the Monte-Carloing over the calibration error. I am curious to see what would be the gain in terms of computation time.} 
We proceed as follows. For each value of the calibration error uncertainty, $\sigma_C$, we generate a large number ($2048$ in this case) of realisations of a data set containing the same GB source plus calibration error with parameters drawn randomly from $\mathcal{N}(0,\sigma_C^2)$. For each realisation, we find the peak of the posterior numerically and hence evaluate the bias in the recovered amplitude and phase of the GW. The distribution of biases over the realisations is Gaussian, with zero mean (because the data generating process and likelihood are consistent), and standard deviation $\sigma_{\mathcal{A}}$ for the GW amplitude and $\sigma_{\phi_0}$ for the GW phase. If the Fisher matrix is valid, these standard deviations should coincide with the Fisher matrix prediction. 

The results from this procedure as shown as open circles in Figure~\ref{fig:GB_ana}. While the agreement is not perfect, it is within a factor of 2 for all choices of $\sigma_C$, indicating the Fisher matrix is providing a very reliable guide to the impact of calibration uncertainty, at a much smaller computational cost than MCMC.

\subsection{Stellar-origin black hole binary semi-analytical confirmation}
In this section we now verify the Fisher matrix for SBHBs. This cannot be done fully analytically, but it is possible to do it without computing numerical derivatives. For an SBHB, the noiseless TDI combinations (here $X$) used to describe the observed gravitational wave can be represented as:
\begin{equation}
\tilde{d}(f) \equiv X(f,\theta,\mu) = C(f,\mu) R_{X}(f,\theta) \mathcal{A}(f,\theta) e^{i \Phi(t)(f,\theta)}
\end{equation}
where $\theta$ are the waveform parameters, $\mu$ the calibration error parameters, $C$ is the calibration error function, $R_X$ is the LISA transfer function for the TDI $X$ combination, $\mathcal{A}$ is the waveform amplitude and $\Phi(t)$ is the waveform phase defined in equation (\ref{eq:GB_WF}). For this source type the LISA transfer function $R_{X}$ depends on all the waveform parameters.

\begin{figure*}
    \centering
    \includegraphics[width=\textwidth]{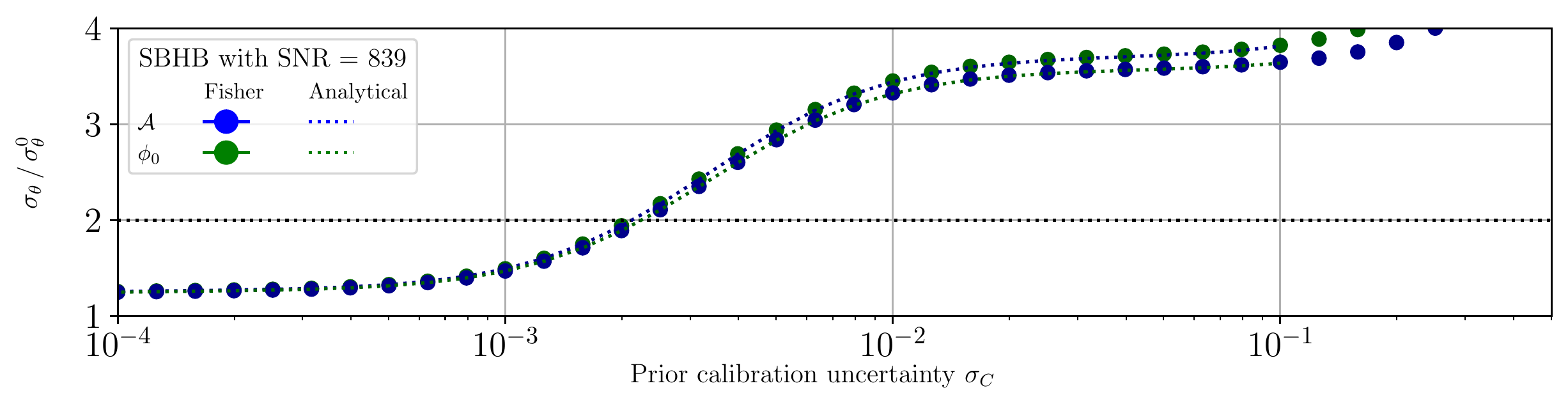}
    \caption{As Figure~\ref{fig:GB_ana}, but now for the SBHB source. The dotted line shows the result of the semi-analytic Fisher matrix evaluation described in this appendix, while the open circles show the results from full numerical evaluation of the Fisher matrix.}
    \label{fig:SBHB_steroid_ana}
\end{figure*}

As shown in the main section of the paper, the SNR is very low for this kind of source. In this appendix, we will artificially increase the SNR (by decreasing the noise power spectral density) in order to reveal the impact of the calibration error. As seen in Figure~\ref{fig:SBHB_steroid_ana}, the initial phase and the distance are mostly affected by the calibration error. Unlike GBs, SBHBs have a broad spectral signature that does not allow the integral over frequency to be approximated by evaluating the integrand at the central frequency, so we still have to evaluate the elements via a numerical integration. However, we can directly write down the derivatives that appear in those integrals. Since the effect on the initial phase is equivalent to that in the case of GBs (subsection \ref{subsec:GB_ana}), we will focus on the chirp mass $\mathcal{M}_c$ and the symmetric mass ratio $\eta$ in this subsection. The generic derivative of the TDI combination $X$ with respect to a parameter $\theta$ is given by :
\begin{equation}
\begin{aligned}
\frac{ \mathrm{d} X(f,\theta,\mu)}{\mathrm{d} \theta} =  \left(\frac{1}{R_x(f,\theta)} \frac{\partial R_x(f,\theta)}{\partial \theta} + \right. \\ 
\left. \frac{1}{\mathcal{A}(f,\theta)} \frac{\partial \mathcal{A}(f,\theta)}{\partial \theta} + i \frac{\partial \Phi(f,\theta)}{\partial \theta}\right) X(f,\theta,\mu)
\end{aligned}
\end{equation}

For the chirp mass $\mathcal{M}_c$ and symmetric mass ratio $\eta$, the transfer function $R_x$ and amplitude $\mathcal{A}$ derivatives can be neglected compared to the phase $\Phi$ derivative :
\begin{equation}
\begin{aligned}
\frac{ \mathrm{d} X(f,\theta,\mu)}{\mathrm{d} \mathcal{M}_c} \simeq  i \frac{\partial \Phi(f,\theta)}{\partial \mathcal{M}_c} X(f,\theta,\mu) , \\
\frac{ \mathrm{d} X(f,\theta,\mu)}{\mathrm{d} \eta} \simeq  i \frac{\partial \Phi(f,\theta)}{\partial \eta} X(f,\theta,\mu)
\end{aligned}
\end{equation}
where the partial derivatives of the phase are analytical functions that can be derived from the IMRPhenomD formulae. The derivatives with respect to the calibration error parameters are the same as for GBs (equation \ref{eq:GB_derivative}). Evaluating the integrals of products of these various derivative expressions over frequency we can obtain an alternative semi-analytic computation of the Fisher matrix. This is compared to the fully numerical results in Figure~\ref{fig:SBHB_steroid_ana}. We can see that the two sets of results are in perfect agreement.

%Unlike the case of GBs, the spectral width of SBHBs is too large to be able to replace the weighted inner cross product by the SNR of the source and therefore the integral must be calculated numerically. With this assumption, the Fisher matrix elements including the chirp mass, symmetric mass ratio and calibration error parameters are thus semi-analytically calculated.

\end{appendix}

\bibliography{calib_bib.bib}% Produces the bibliography via BibTeX.

\end{document}